\documentclass[13pt]{article}

\usepackage{arxiv}

\usepackage[utf8]{inputenc} 
\usepackage[T1]{fontenc}    
\usepackage{hyperref}       
\usepackage{url}            
\usepackage{booktabs}       
\usepackage{amsfonts}       
\usepackage{nicefrac}       
\usepackage{microtype}      
\usepackage{graphicx}
\usepackage{doi}
\usepackage{setspace} 
\usepackage{amsmath}
\usepackage{float} 
\usepackage{subfigure} 

\title{Two-stage Contextual Transformer-based Convolutional Neural Network for Airway Extraction from CT Images
\thanks{This work was supported by the National Natural Science Foundation of China (82072008), Natural Science Foundation of Liaoning Province (2021-YGJC-21), Key R\&D Program Guidance Projects in Liaoning Province (2019JH8/10300051), Hong Kong Research Grants Council (RGC) Collaborative Research Fund (C4026-21G), Hong Kong General Research Fund (RGCC-GRF 14216022) and the Fundamental Research Funds for the Central Universities (N2124006-3, N2224001-10).}} 

\author{
\vspace{0.5em}
{\hspace{1mm}Yanan Wu$^{1,2,4}$}
{\hspace{1mm}Shuiqing Zhao$^{1}$}
{\hspace{1mm}Shouliang Qi$^{1,2}$*}
{\hspace{1mm}Jie Feng$^{3}$} 
{\hspace{1mm}Haowen Pan$^{1}$}
{\hspace{1mm}Runsheng Chang$^{1}$} \\ \vspace{0.5em}
{\hspace{1mm}\textbf{Long Bai}$^{4}$}
{\hspace{1mm}\textbf{Mengqi Li}$^{5}$}
{\hspace{1mm}\textbf{Shuyue Xia}$^{6}$}
{\hspace{1mm}\textbf{Wei Qian}$^{1}$}
{\hspace{1mm}\textbf{Hongliang Ren}$^{4}$*}
\\
\vspace{0.5em}
 $^{1}$College of Medicine and Biological Information Engineering, Northeastern University, Shenyang, China\\
\vspace{0.5em}
 $^{2}$Key Laboratory of Intelligent Computing in Medical Image, Northeastern University, Shenyang, China\\
 \vspace{0.5em}
$^{3}$School of Chemical Equipment, Shenyang University of Technology, Shenyang, China \\
\vspace{0.5em}
$^{4}$Department of Electrical Engineering, The Chinese University of Hong Kong, Hong Kong, China \\
\vspace{0.5em}
$^{5}$Department of Respiratory, Second Affiliated Hospital of Dalian Medical University, Dalian, China \\
\vspace{0.5em}
$^{6}$Respiratory Department, Central Hospital Affiliated to Shenyang Medical College, Shenyang, China\\
\vspace{0.5em}
\texttt{*corresponding author:} 
\texttt{qisl@bmie.neu.edu.cn}, \texttt{hren@cuhk.edu.hk} \\
}

\date{}

\renewcommand{\shorttitle}{}
\hypersetup{
pdftitle={A Two-stage Contextual Transformer-based Convolutional Neural Network for Airway Extraction from CT Images: Methods and Applications},
pdfsubject={q-bio.NC, q-bio.QM},
pdfauthor={Yanan Wu},
pdfkeywords={Contextual Transformer, image segmentation, convolution neural network, computed tomography},
}
\makeatletter
\def\thanks#1{\protected@xdef\@thanks{\@thanks
        \protect\footnotetext{#1}}}
\makeatother
\begin{document}
\maketitle
\setlength{\baselineskip}{15pt}

\begin{abstract}
	Accurate airway extraction from computed tomography (CT) images is a critical step for planning navigation bronchoscopy and quantitative assessment of airway-related chronic obstructive pulmonary disease (COPD). The existing methods are challenging to sufficiently segment the airway, especially the high-generation airway, with the constraint of the limited label and cannot meet the clinical use in COPD. We propose a novel two-stage 3D contextual transformer-based U-Net for airway segmentation using CT images. The method consists of two stages, performing initial and refined airway segmentation. The two-stage model shares the same subnetwork with different airway masks as input. Contextual transformer block is performed both in the encoder and decoder path of the subnetwork to finish high-quality airway segmentation effectively. In the first stage, the total airway mask and CT images are provided to the subnetwork, and the intrapulmonary airway mask and corresponding CT scans to the subnetwork in the second stage. Then the predictions of the two-stage method are merged as the final prediction. Extensive experiments were performed on in-house and multiple public datasets. Quantitative and qualitative analysis demonstrate that our proposed method extracted much more branches and lengths of the tree while accomplishing state-of-the-art airway segmentation performance. The code is available at \href{https://github.com/zhaozsq/airway\_segmentation}{https://github.com/zhaozsq/airway\_segmentation}.
\end{abstract}


\section{Introduction}
Computed tomography (CT) plays an increasingly important role as a predominant medical imaging tool in diagnosing and assessing human diseases, especially lung disease. Accurate airway segmentation in CT images plays a crucial role in the planning of navigation bronchoscopy \cite{benn2021robotic} and the assessment of airway-related chronic obstructive pulmonary disease (COPD) \cite{tanabe2021central,lu2021necroptosis}. A lung biopsy can sometimes be conducted if a nodule is seen on the CT scans to identify whether it is benign. Navigation bronchoscopy through the airway tree is the most effective lung biopsy technique for reaching peripheral pulmonary lesions \cite{ishiwata2020bronchoscopic} or the target bronchioles \cite{kemp2020navigation,asano2014virtual,edell2010navigational}. Additionally, the localization of the endoscopic tips is a key component of the present bronchoscopic roadmap  navigation systems \cite{shen2019context,mehta2018evolutional,higgins2015multimodal}, which necessitates using a superior 3D airway model.

Variable degrees of emphysema and airway disease can be found in COPD, a heterogeneous disorder \cite{halpin2021global}. Clinically, these illnesses are frequently grouped and measured by the severity of airflow limitation. CT scanning can visually differentiate airway remodeling \cite{hirota2013mechanisms,ding2016measuring} and emphysema \cite{goddard1982computed}, which are underlying pathological subtypes. It has demonstrated that the airway wall thickness has been a biomarker assessing COPD \cite{mets2013diagnosis} and corresponds with airflow obstruction \cite{sasaki2014ratios,lutey2013accurate}, and is easily measurable on inspiratory lung screening CT.
\begin{figure}[H] 
    \vspace{-1em}
    \centering 
    \includegraphics[width=0.7\textwidth]{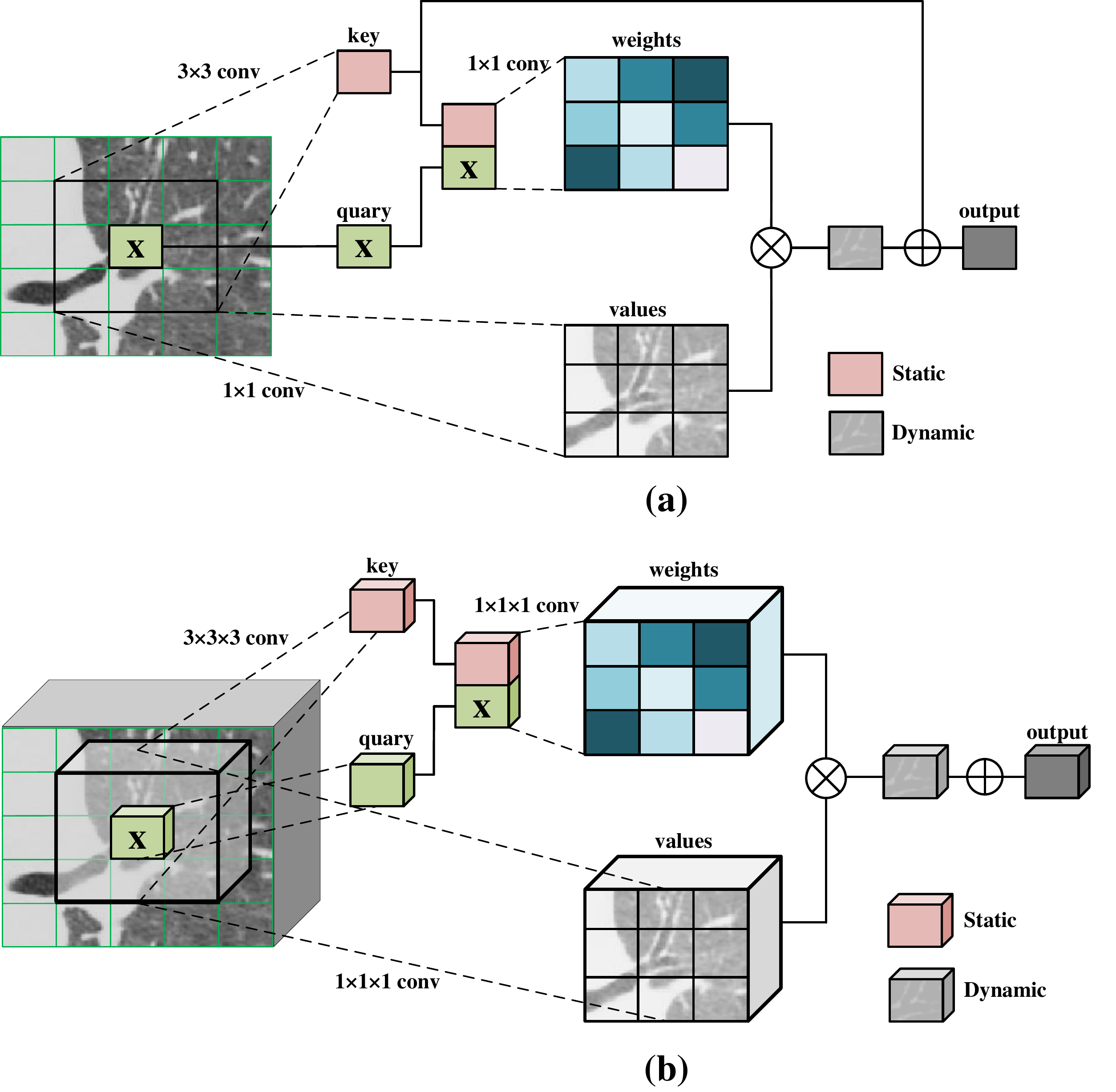} 
    \caption{Comparison between 2D and 3D Contextual Transformer block. (a) 2D Contextual Transformer block (b) 3D Contextual Transformer block.} 
    \label{fig1} 
\end{figure}
Over the past few years, deep learning has gained popularity in computer vision and image processing, including natural and medical images. They have shown particular strengths in extracting the best knowledge from the input data and taking full advantage of it without application-specific rules or manual-design features \cite{litjens2017survey,lecun2015deep}. The four significant aspects of using DL in medical imaging are classification, object localization, object detection, and segmentation. The segmentation of organs \cite{luo2022word,ma2022fast}, other substructures \cite{chen2021diverse,momin2022mutual}, and lesions \cite{feng2022bla,wu2021elnet,chen2021effective} in multi-modality medical images enables additional quantitative analysis and uses them as the initial step in computer-aided detection (CAD) systems. Patches of an input image or the entire image are typically the two different input sources used for image segmentation in medical imaging based on the deep learning approach. 

The vision transformer (ViT) \cite{dosovitskiy2020image}, which is intended to learn different visual representations from convolutional neural networks (CNNs), has recently demonstrated comparable or even better performance than CNNs. The transformer was at first introduced in natural language processing (NLP) \cite{devlin2018bert} and was based on the self-attention mechanism \cite{vaswani2017attention}. Recently, ViT has been effectively and competitively applied to computer vision (CV), including semantic segmentation \cite{valanarasu2021medical}, image classification \cite{wu2021vision,zhao2022cot}, and image recognition \cite{dosovitskiy2020image}. ViT can acquire local and global representations and preserves more spatial information than CNNs \cite{raghu2021vision}.

In this work, we adopt and modify the Transformer-style block in an elegant way from \cite{li2022contextual}. In \cite{li2022contextual}, Contextual Transformer (CoT) block is first proposed for image recognition, object detection, and instance segmentation in the natural image field. The main intention of the CoT block is proposed to exploit the rich information of context over 2D feature map. As shown in Fig. \ref{fig1}(a), the key tasks are the static and dynamic contextual representation of inputs. First, a 3×3 convolution layer is employed to contextualize the representation of keys. The acquired contextualized feature is used as the static context representation of inputs among local neighbors. Then, the input query and the contextualized key feature are concatenated. Two 1×1 convolution layers are followed with the concatenation operation to generate the attention matrix. CoT takes advantage of the relationships between each query and each key, much like self-attention learning. The generated attention matrix is then used to combine all of the input data, creating the dynamic context. The static and dynamic contexts are finally combined as the output. In medical image analysis, especially in CT images, the 3D volumetric data can provide large-range information and global queues for tubular structure segmentation, such as airway and pulmonary vessels. Therefore, we first utilize and modify the vanilla CoT block for 3D volumetric data. As described in Fig. \ref{fig1}(b), the 2D convolution layers (1×1, 3×3) are replaced by the 3D convolutions (1×1×1, 3×3×3). 
\begin{figure}[H] 
    \vspace{-1em}
    \centering 
    \includegraphics[width=0.9\textwidth]{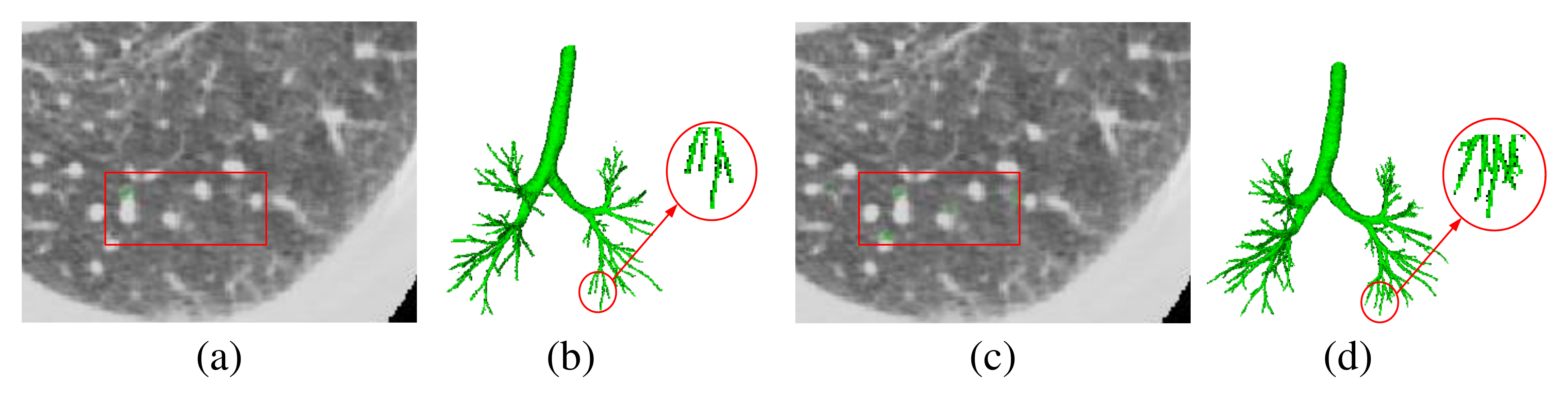} 
    \caption{(a) Axial view CT image patch with the label showing the mislabeled airway (red box); (b) airway tree labeled by experienced radiologists; (c) Axial view CT image patch with prediction of proposed method showing more segmentation of airway; (d) airway tree prediction by proposed method.} 
    \label{fig2} 
\end{figure}
Although advanced techniques and methods are developed, accurate and robust airway segmentation remains challenging from two aspects, as noted in \cite{zheng2021alleviating,wang2022naviairway}. The first type of class imbalance is significantly fewer airway voxels than background voxels in CT images, referred to as inter-class imbalance. The performance will be limited by false positives and breakages of small airway branches if the inter-class imbalance is not considered in a segmentation pipeline. In this study, the contextual transformer is used to enhance the airway points segmentation performance in the presence of significant local inter-class imbalance. The CoT block is proposed to exploit the rich information of context over 2D feature map using modified input keys \cite{li2022contextual}. In this paper, we extend 2D CoT module to 3D architecture and add it to the encoder and decoder part of 3D U-Net architecture. The other imbalance, the intra-class imbalance, pertains to the differences between the low-generation airway and distal segmental bronchi. According to the morphological complexity of airway in CT images, the bulk of the volume is made up of large airways, and for data-driven (deep learning) algorithms, this uneven distribution influences how well the peripheral bronchi are segmented. And in the high generation of the airway, labeling them is also difficult for the expert. As described in Fig. \ref{fig2}, a few labels of small airway segments were missing in annotations. Nevertheless, the results by proposed method were generally still of good quality. Especially, some airways are mislabeled by radiologists in Fig. \ref{fig2}(a). A two-stage U-Net-like network is proposed for accurate and robust airway segmentation and to compensate for the manual labeling shortage (Fig. \ref{fig2}(c)).

In this study, we proposed a two-stage contextual transformer-based modified 3D U-Net model to address the above two imbalances in airway segmentation using CT images. The main contributions of this work are as follows: First, the contextual transformer is introduced and extended to 3D architecture better to exploit the richness of context in the CT images. Second, a two-stage network is proposed for airway segmentation with peripheral airway refinement. The specialized small airway model (the second stage model) are served as supplementary knowledge to further refine the airway segmentation. Third, multi-dataset and multi-disease applications were brought into the study. And it demonstrated that the proposed airway segmentation algorithm is robust and usable when extending to other datasets.

\section{Related work}
\textbf{Traditional airway segmentation methods} Many traditional airway segmentations have been proposed, including region growing \cite{fabijanska2009two,pinho2009robust}, thresholding \cite{shi2006upper,aykac2003segmentation}, template matching \cite{born2009three}, and gradient vector flow \cite{bauer2009segmentation}, fuzzy connectivity \cite{tschirren2005segmentation,tschirren2005intrathoracic}, and hybrid method \cite{kiraly2002three,meng2017automatic}. Additionally, EXACT’09 Challenge \cite{lo2012extraction} proposed fifteen algorithms in which most of the method were region growing and thresholding.

\textbf{Machine learning and deep learning airway segmentation methods} Following development of deep learning in computer vision, deep learning, especially CNNs have also garnered attention in pulmonary airway segmentation. Some studies employed CNN with specific post-processing methods, for example, U-Net \cite{ronneberger2015u}. The post-processing methods consists of region growing and skeletonization \cite{nadeem2020ct}, image boundary \cite{garcia2018automatic}, and freeze-and-grow propagation \cite{jin20173d}. Moreover, other studies are dedicated to modifying the U-Net model. Wang et al. proposed a network which applied U-Net architecture as baseline and processed the 3D CT scan slice-by-slice, and designed a coined radial distance loss for tubular structures \cite{wang2019tubular}. Juarez and his colleagues combined the U-Net and graph neural network (GNN) \cite{garcia2019joint}. Moreover, Selvan et al. introduced the GNN method and refined it with graph refinement module \cite{selvan2020graph}. In recent years, more strategies attached to CNN are proposed for airway tasks, including attention distillation and feature recalibration modules \cite{qin2021learning}, group supervision \cite{zheng2021alleviating}, a fusion of attention mechanism and fuzzy theory \cite{nan2022fuzzy}, and a semi-supervised learning framework \cite{wang2022naviairway}.

\textbf{Transformer method in image segmentation} With the success of transformer in NLP, the transformer-based method has been developed in vision tasks, including medical image segmentation. TransUNet \cite{chen2021transunet} was  proposed with a combination of CNN and Transformer architecture, which intended to fuse the spatial feature from CNN and the global context generated from Transformers for medical image segmentation. Swin-transformer employs a modified self-attention block with the shifted window to construct a hierarchical representation and achieved strong performance on semantic segmentation \cite{liu2021swin}. SegFormer consists of a novel positional-encoding-free module. Moreover, its encoder is a hierarchical Transformer and the decoder is a lightweight multilayer perceptron \cite{xie2021segformer}. SEgmentation Transformer (SETr) \cite{zheng2021rethinking} exploits the transformer as a baseline and the novel encoder with fully attention by sequentializing images. The proposed method achieved SOTA performance for image segmentation using transformer-like architecture. Lin et al. \cite{lin2022ds} proposed a U-like architecture that employed swin transformer block in the encoder and decoder paths. The main contribution of Lin’s model was the dual-scale encoding module.

\section{Datasets and methods}

\subsection{Datasets}
	\paragraph{In-house dataset} this dataset includes 23 CT scans (slice thickness=1mm) for patients with COPD from two hospitals and 20 CT scans for training with annotations from EXACT’09 \cite{lo2012extraction}. Among the 23 cases, 12 are from Central Hospital Affiliated to Shenyang Medical College, and 11 are from the Second Hospital of Dalian Medical University. All participants provided informed consent in accordance with the Declaration of Helsinki (2000), and the study was approved by the two hospitals. The airway masks for 23 CT scans are annotated in a semi-automatic way. First, the ‘deep airway segmentation’ module of Mimics software (Materialise Corp, Belgium) is employed to initially segment the mask of the airway. The airway annotations are then manually modified and improved by two skilled radiologists.
	\paragraph{EXACT’09 challenge} 20 CT scans with labels and 20 CT scans without available annotation are included \cite{lo2012extraction}. This dataset contains patients with illnesses ranging from normal to severe pulmonary lung disorders. The challenge organizers calculated the test metrics in this dataset for fair comparison experiments.
	\paragraph{ISICDM 2021 Challenge dataset} ISICDM 2021 includes 12 cases of non-contrast CT scans \cite{tan2021analysis,tan2021automated,tan2022segmentation}. The thickness of CT scans is between 0.625 and 1.5 mm, and the number of slices was between 204 and 577. The matrix size of all-CT images was 512×512. The original images are in DICOM format, and the relevant airway masks are in JPG format. 
	\paragraph{Binary airway segmentation (BAS) dataset} The dataset consists of 90 CT scans in total, of which 20 instances are from EXACT'09, and 70 cases are from the image collection of the Lung Image Database Consortium (LIDC-IDRI). One thousand eighteen thoracic CT images for diagnostic and screening purposes for lung cancer are included in the LIDC-IDRI dataset \cite{armato2011lung}. 70 examples (slice thickness 0.625mm) out of the 1018 cases were chosen and annotated by specialists \cite{zheng2021alleviating}. According to this work \cite{zheng2021alleviating}, we divided the dataset into three sets: a training set of 50 cases, a validation set of 20, and a test set of 20 cases. 
	\paragraph{Airway Tree Modeling 2022 (ATM22) dataset} The challenge dataset contains 500 computed tomography (CT) scans (300 for training, 50 for external validation, and 150 for testing), which were collected from the public LIDC-IDRI dataset and the Shanghai Chest hospital \cite{zheng2021alleviating,qin2019airwaynet,yu2022break,zhang2021fda}. The CT images are initially preprocessed using robust deep-learning models and an ensemble technique to obtain the preliminary segmentation results. Three radiologists with the combined expertise of more than five years then painstakingly outline and double-check these results to obtain the final refined airway tree structure. One scan was removed due to incorrect labeling, as requested by the challenge organizers. Therefore, this paper used the official training set (299 CT scans) for model training. The metrics were calculated by the challenge organizers.
	\paragraph{COPD dataset} This dataset contains 420 cases from Central Hospital Affiliated with Shenyang Medical College and the Second Hospital of Dalian Medical University. According to the Global Initiative for Chronic Obstructive Lung Disease (GOLD), the COPD patient was determined with FEV1/FVC < 0.7. Meanwhile, the definition of COPD severity was as follows, mild COPD (stage I), FEV1$\geq$ 80\% predicted; moderate COPD (stage II), 50 \% $\leq$ FEV1<80 \% predicted; severe COPD (stage III), 30 \% $\leq$ FEV1<50 \%  predicted; and very severe COPD (stage IV), FEV1<30 \% predicted. The number of cases in each GOLD stage is stage I: 80, Stage II: 77, Stage III: 116, and Stage IV: 70. Moreover, the number of healthy control is 77, and slice thickness is 1mm.
	\paragraph{COVID-19 dataset} this dataset consists of 96 CT scans (0.625mm$\leq$ slice thickness$\leq $1.25mm) of COVID\-19 patients selected from the RSNA International COVID-19 Open Radiology Database (RICORD) dataset. The RICORD dataset consists of 240 thoracic CT scans and 1000 chest radiographs from four sites in different countries.
\begin{figure}[H] 
    \vspace{-1em}
    \centering 
    \includegraphics[width=0.9\textwidth]{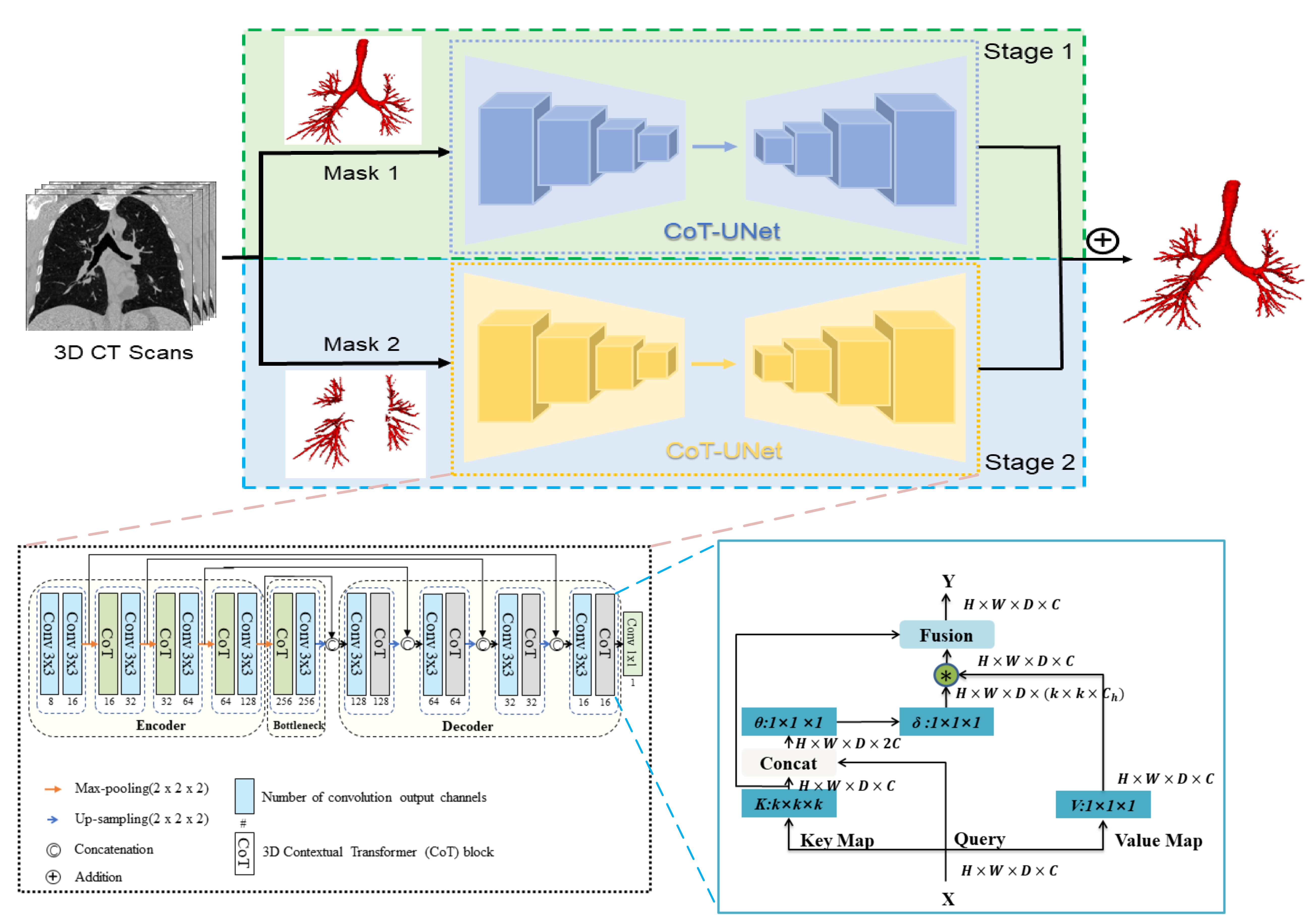} 
    \caption{Workflow of the proposed method for airway segmentation.} 
    \label{fig3} 
\end{figure}
\subsection{Overview of the proposed method and architecture}
The proposed method is described in Fig. \ref{fig3}. The method consists of two stages performing initial and refined segmentation, respectively. In stage 1, the total airway mask and 3D computed tomography (CT) scans are fed to the proposed network, and the intrapulmonary airway mask and 3D CT scans are fed to the model in stage 2. Then the results of the two stages are merged as the final prediction.
In each stage, the same subnetwork is utilized in the method as described in Fig. \ref{fig3}. The model is derived from 3D U-Net \cite{cciccek20163d} with the encoder-decoder structure as the basis and modified by replacing some 3x3x3 convolution layers with the contextual transformer (CoT) module. The encoder-side employs four pooling layers with five resolution scales for enlarging the receptive field. At each scale, except for the first scale in encoder path, both encoder and decoder have a convolutional layer (kernel size 3 × 3× 3) and a 3D CoT module, followed by instance normalization and Rectified Linear Units (ReLU). A skip connection is applied to concatenate the feature layer between the encoder and decoder.

\subsection{Preprocessing}
In the preprocessing, first, the voxel intensities of the lung CT images were truncated within the window from -1,000 to 600. Second, the CT value is normalized between 0 and 1. Then, lung area segmentation is implemented. After that, we crop the region of interest of the CT images and the corresponding labels for both the training and validation phases.
In lung segmentation, the methodology was first reported in the paper \cite{hofmanninger2020automatic} which examined four widely used deep learning models on diverse datasets. The highest Dice Similarity Coefficient (DSC) was obtained by the U-Net (R-231) with 0.98±0.03. Therefore, the pre-trained U-Net (R-231) model was employed to segment the lung field.
\subsection{Contextual Transformer}
\begin{figure}[H] 
    \vspace{-1em}
    \centering 
    \includegraphics[width=0.9\textwidth]{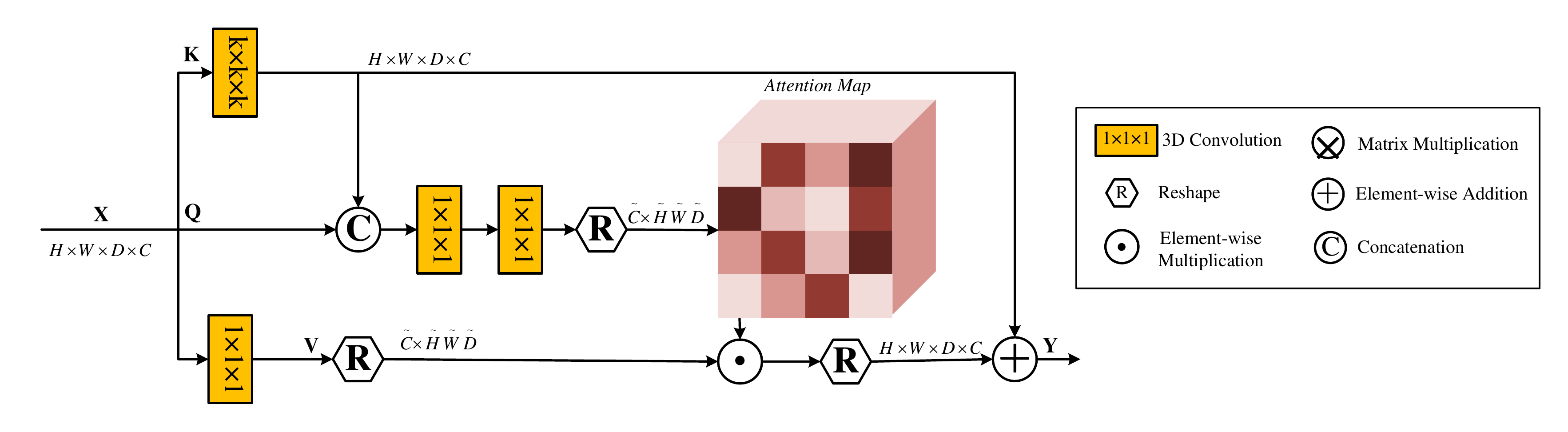} 
    \caption{The architecture of 3D contextual transformer.} 
    \label{fig4} 
\end{figure}
The CoT block is proposed by Li et al. \cite{li2022contextual} to exploit the rich contexts in the input features. The proposed 2D CoT block limits the application capacity in 3D volumetric segmentation. For extending the application of CoT, a new 3D CoT block is constructed in Fig. \ref{fig4}. Formally, the input 3D feature map $X \in \mathbb{R}^{H \times W \times D \times C}$ with resolution (H,W,D) and channel number C is transformed into keys K, queries Q, and values V, respectively. 

\begin{equation}
	V=\mathrm{XW}_{v}
\end{equation}
where $\mathrm{W}_{v}$ is the embedding matrix.
Then, the static contextual information $K^{1} \in \mathbb{R}^{H \times W \times D \times C}$ of input X is produced by employing k×k×k convolution over all the neighbor keys for contextualizing each key K representation. After that, contextualized keys $K^{1}$ and queries Q are concatenated, followed by two consecutive 1×1×1 convolutions the attention matrix A. 
\begin{equation}
	A=\left[K^{1}, Q\right] W_{\theta} W_{\delta}
\end{equation}
where $W_{\theta}$ and $W_{\delta}$  are the learnt parameters.
Next, the multiplication of the feature map $K^{2}$ and values V is used as dynamic contextual representation.
\begin{equation}
	K^{2}=V * A
\end{equation}
Finally, the output of the proposed CoT block is measured as the fusion of the static contextual feature map $K^{1}$ and dynamic contextual feature map $K^{2}$.
\subsection{Focal and Dice Loss}
In this study, it is typical for the relevant airways to occupy only very small spaces in each slice of CT images. Therefore, to solve the class imbalance, the combination of Dice loss \cite{milletari2016v} and Focal loss \cite{lin2017focal} is employed for training the network. The model training loss is defined as follows.
\begin{equation}
\text { Total }_{\text {loss }}=D_{\text {loss }}+F_{\text {loss }}
\end{equation}
For Dice loss,
\begin{equation}
    D_{\text {loss }}=1-\frac{2 \times \sum_{i=1}^{N} p_{i} l_{i}}{\sum_{i=1}^{N}\left(p_{i}+l_{i}\right)}
\end{equation}

Where $p_{i}$ is the binary airway prediction and $l_{i}$ is the airway label.

For Focal loss,
\begin{equation}
F_{\text {loss }}=-\alpha\left(1-p_{\mathrm{t}}\right)^{\gamma} \log \left(p_{\mathrm{t}}\right)
\end{equation}

Where $p_{\mathrm{t}}$ is probability of predicting the airway. $\alpha$ and $\gamma$ control the class weights and degree of down-weighting of easy-to-classify pixels, respectively. In this paper, $\alpha=$0.25 and $\gamma=$2.

\subsection{Experimental Setup}
The number of epochs is 50. The initial learning rate is 0.003. The batch size is 2. Adam optimizer is used in the training process. Moreover, data augmentation was utilized in training via the flip and jitter. And early stopping was adopted to alleviate the overfitting when  validation accuracy did not increase within five epochs.
All experiment was conducted on a workstation with a central processing unit of Intel(R) Xeon(R) Silver 4114 CPU, 128GB RAM, and four NVIDIA GeForce RTX 2080Ti Graphical Processing Unit (with 11GB memory). The popular PyTorch framework is utilized and the code is written in Python.

\subsection{Evaluation Metric of Airway Segmentation}
After the largest connected component of the output was implemented, five metrics were used to evaluate the performance of the proposed model: Branches detected (BD); Tree-length detected (TD); True positive rate (TPR); False positive rate (FPR); Dice similarity coefficient (DSC). The definitions of metrics are referred to \cite{qin2021learning}.
The calculation of BD, TD, TPR, and FPR only considers the lung region to indicate the model's capacity to detect peripheral airways. The trachea is nonetheless used in DSC calculations since it determines the overall effectiveness of segmentation.

\section{Experiments and results}

\subsection{Performance of the proposed model and models with ablations}
\begin{figure}[H] 
    \centering 
    \includegraphics[width=0.6\textwidth]{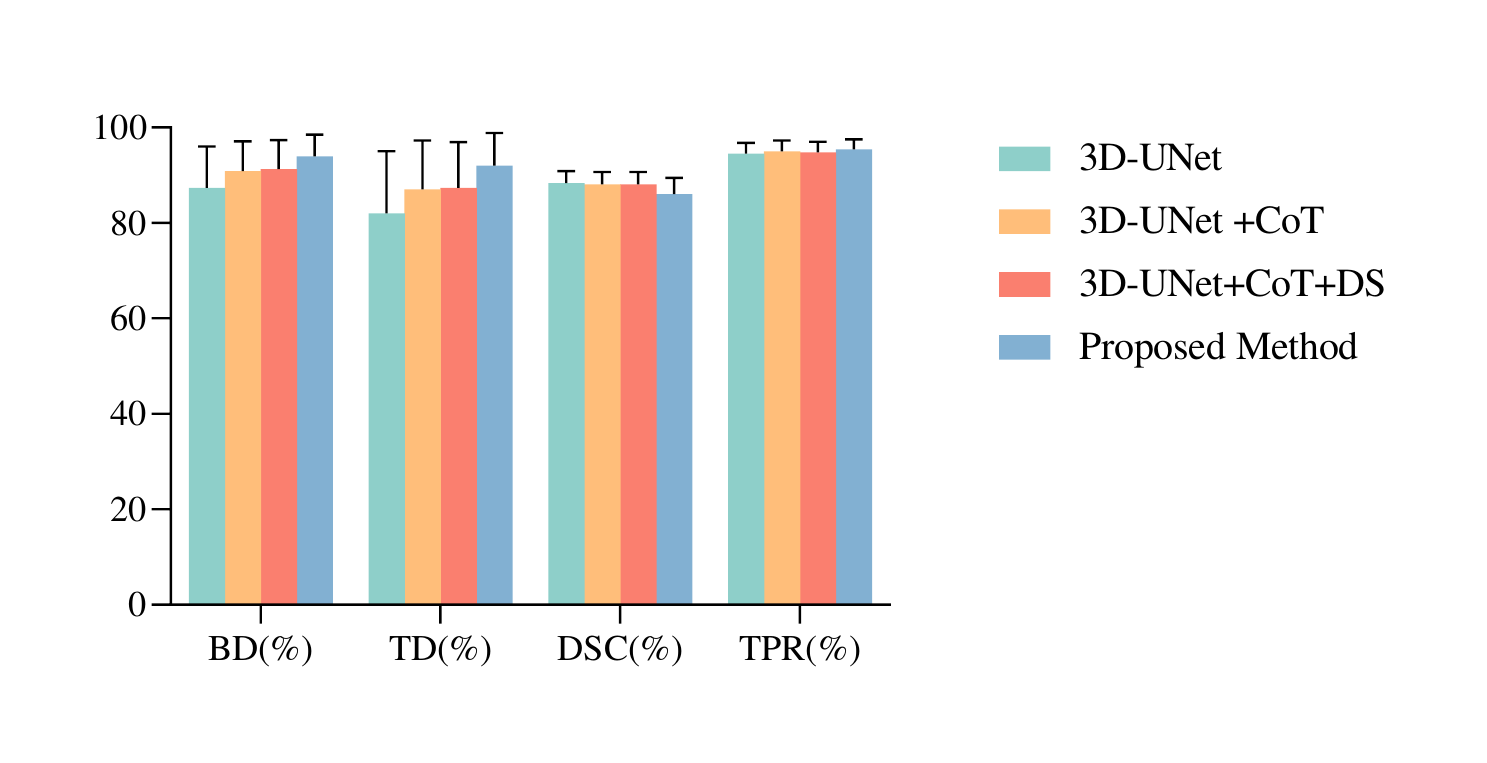} 
    \caption{The segmentation metric of proposed method and ablation experiments on ISICDM 2021 dataset.} 
    \label{fig5} 
\end{figure}

\begin{table}[H]
\caption{Performance of the proposed model and models with ablations on the ISICDM 2021 dataset. “DS” denotes deep supervision strategy.}
\centering
\begin{tabular}{llllll}
\hline
\textit{Model}  & \textit{TD (\%)} & \textit{BD (\%)} & \textit{DSC (\%)} & \textit{TPR (\%)} & \textit{FPR (\%)} \\ \hline
3D-UNet         & 87.41±8.63       & 82.08±12.95      & 88.39±2.49        & 94.51± 2.28       & 0.118±0.029       \\
3D-UNet+CoT     & 90.87±6.23       & 87.03±10.29      & 88.14±2.54        & 95.06±2.19        & 0.133±0.032       \\
3D-UNet+CoT+DS  & 91.33±6.02       & 87.37± 9.61      & 88.10±2.63        & 94.81±2.20        & 0.136±0.025       \\
Proposed Method & 93.96±4.54       & 92.06±6.83       & 86.05±3.37        & 95.42±2.10        & 0.193±0.045       \\ \hline
\end{tabular}
\label{table1}
\end{table}
On the ISICDM2021 dataset, our proposed method, which adopts 3D U-Net with CoT module and deep supervision strategy, achieves high performance with 93.96\% TD and 92.06\% BD. To analyze the important modules in our proposed model, the CoT module, deep supervision strategy, and the two-stage network are selected at the vanilla 3D U-Net which is as the baseline. For the ablation study, the in-house airway and EXACT’09 training datasets were utilized for training and validating the model. The ISICDM 2021 dataset was employed as the test set. The quantitative results are shown in Table \ref{table1} and Fig. \ref{fig5}. The baseline model (3D-UNet) achieves competitive performance with an 87.41\% TD, 82.08\% BD, 88.39\% DSC, 94.51\% TPR, and 0.118\% FPR. Replacing some convolutional layers with the CoT layer in the baseline model shows a considerable improvement in segmentation performance on TD (3.46\%) and BD (4.95\%) scores. By adopting a deep supervision strategy with 3D-CoT-U-Net, the model also achieves a slightly better improvement with a 0.46\% increment of TD and 0.34\% of BD. By integrating the above strategy and adopting the two-stage network, the largest improvement is witnessed with 93.96\% TD and 92.06\% BD. However, a reduction is observed in the DSC score (86.05\%).
\begin{figure}[H] 
    \vspace{-1em}
    \centering 
    \includegraphics[width=0.8\textwidth]{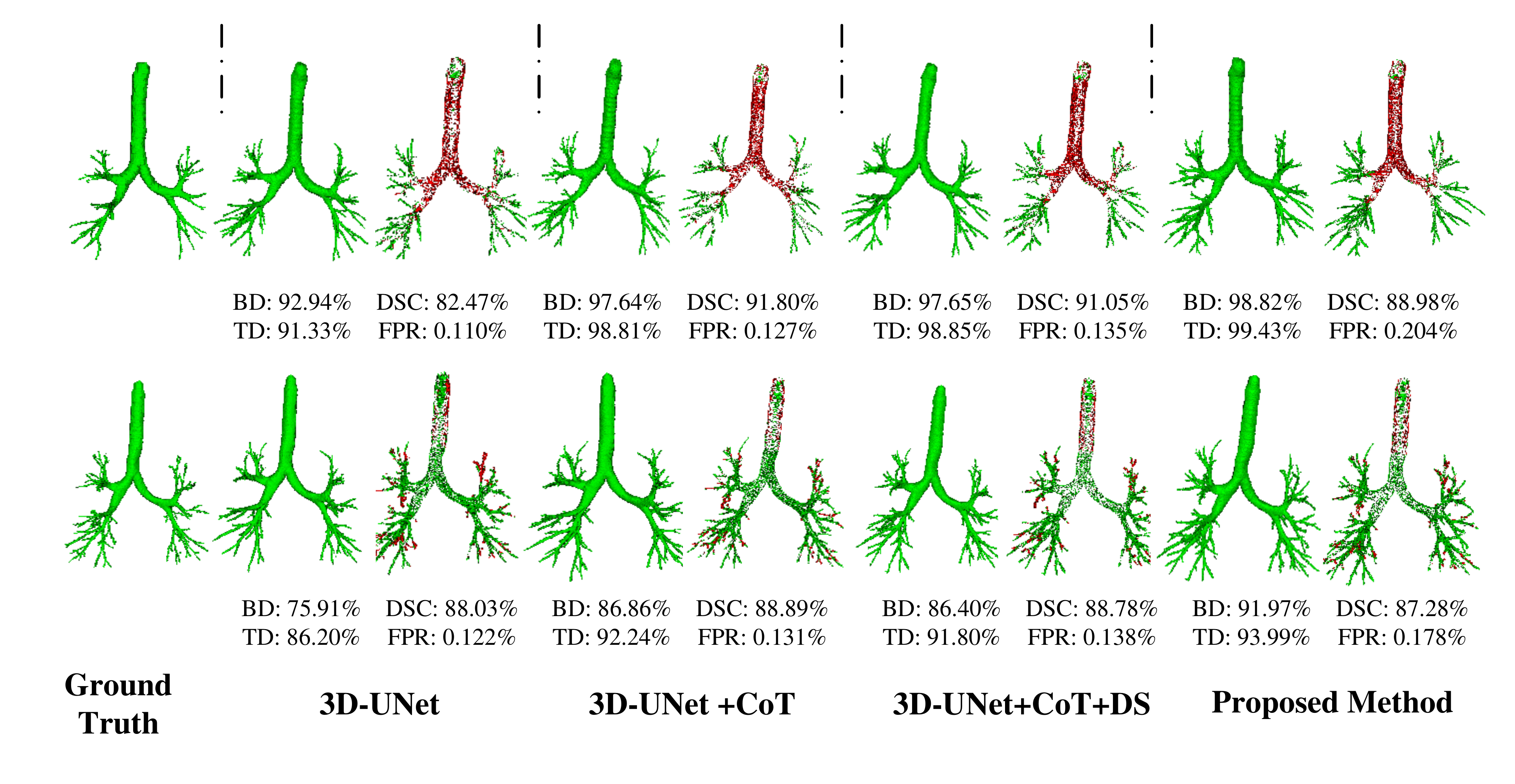} 
    \caption{The visualization of segmentation results of the proposed method and the ablation experiment. For each method, the left figure is the segmentation result, while the right illustrates the false positives in green and the false negatives in red.} 
    \label{fig6} 
\end{figure}
Fig. \ref{fig6} visualizes the segmentation results and ablation experiments. The CoT block, the deep supervision, and a two-stage strategy were tuned on the baseline 3D U-Net model. In the baseline model, the large and some small airways from the baseline results can roughly appear. However, the distal branches are represented more seriously. Besides, with a CoT block, great improvements were observed with TD (from 92.94\% to 97.64\%, from 75.91\% to 86.86\%) and BD (from 91.33\% to 98.81\%, from 86.20\% to 92.24\%). Moreover, adopting the deep supervision strategy achieved the comparative TD, BD, and DSC, but a relatively higher FPR. From Fig. \ref{fig6}, over-segmentation situations can be observed in the distal airway and these may be the true airways. Under a two-stage network, the method achieved a higher TD, BD, and FPR but a decline in DSC score, leading to more detected peripheral branches, even unannotated by radiologists.
\subsection{Performance comparison on public ISICDM2021 dataset}
\begin{table}[H]
\caption{Performance comparison on the ISICDM 2021 dataset.}
\centering
\begin{tabular}{llllll}
\hline
\textit{Dataset} & \textit{BD (\%)} & \textit{TD (\%)} & \textit{DSC (\%)} & \textit{TPR (\%)} & \textit{FPR (\%)} \\ \hline
AirwayNet \cite{qin2019airwaynet}        & 88.44±7.43       & 83.58±11.82      & 87.88±2.77        & 94.02±3.01        & 0.125±0.032       \\
nn-UNet \cite{isensee2021nnu}          & 90.55±8.35       & 86.83±12.69      & 89.37±2.03        & 95.02±2.34        & 0.109± 0.023      \\
Wingsnet \cite{zheng2021alleviating}         & 91.25±5.76       & 87.25±   9.06    & 88.89±3.04        & 93.61±2.94        & 0.117±0.029       \\
Proposed Method  & 93.96±4.54       & 92.06±6.83       & 86.05±3.37        & 95.42±2.10        & 0.193±0.045       \\ \hline
\end{tabular}
\label{table2}
\end{table}
To further prove the performance of the proposed method, comparison experiments were conducted on the ISICDM2021 public dataset. For the ISICDM2021 dataset, the training dataset is the in-house airway and EXACT’09 training datasets. 
As shown in Table \ref{table2}, the proposed method achieves state-of-the-art (SOTA) performance with better BD, TD, and TPR scores of 93.96\%, 92.06\%, and 95.42\%, but lower DSC (86.05\%) and FPR (0.193\%) ISICDM2021 dataset. WingNet \cite{zheng2021alleviating} achieved competitive results with 91.25\% BD, 87.25\% TD, 88.89\% DSC, 93.61\% TPR, and 0.117\% FPR. The nn-UNet \cite{isensee2021nnu} and Wingsnet \cite{zheng2021alleviating} achieve high DSC of 89.37\% and 88.89\%, and lower FPR scores with 0.109\% and 0.117\%, representing the powerful studying ability with the provided annotation airway. The relative lower performance of BD and TD are observed in AirwayNet \cite{qin2019airwaynet} with 88.44\% and 83.58\%, respectively.
\begin{figure}[H] 
    \vspace{-1em}
    \centering 
    \includegraphics[width=0.9\textwidth]{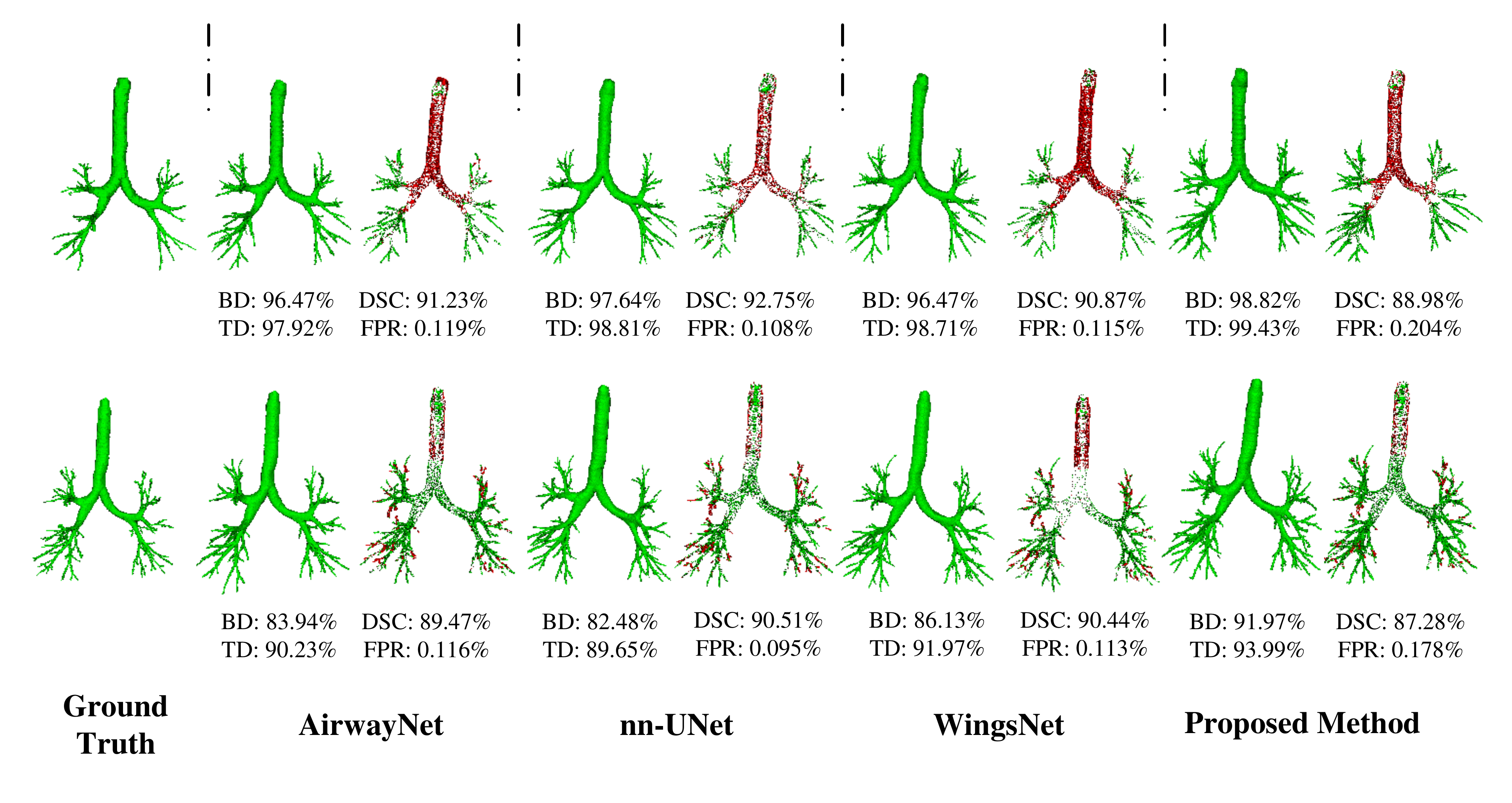} 
    \caption{The segmentation results of the different methods on ISICDM 2021 dataset. For comparative methods, the segmentation result is on the left, while the right describes the false positives in green and the false negatives in red.} 
    \label{fig7} 
\end{figure}
Fig. \ref{fig7} visualizes the comparison with the SOTA method using CNNs. The proposed method achieved the highest TD, BD, and FPR in the two examples. Compared to AirwayNet, nnU-Net, and WingsNet, more peripheral branches were successfully detected by the proposed method, and less discontinuity was observed in the airway prediction regions. Furthermore, nnU-Net performed better on the DSC metric due to its great power in label learning. Compared with the other three methods, our proposed method was prone to detect more thin bronchioles. The two-stage network could contribute to the phenomenon. 

\subsection{Performance comparison on public EXACT’09 dataset}
For the EXACT’09 challenge, the model is trained with the in-house airway and the training dataset of EXACT’09. And we submitted our results on the test dataset to the challenge organizer for evaluation.
\begin{table}[H]
\caption{Performance comparison on EXACT’09 dataset.}
\centering
\begin{tabular}{lllll}
\hline
\textit{Dataset}  & \textit{BD(\%)} & \textit{TD(\%)} & \textit{FPR(\%)} & \textit{Precision(\%)} \\ \hline
Irving et al. \cite{irving20093d}     & 43.5±19.1       & 36.4±17.1       &                  & 98.7±2.9               \\
Pinho et al. \cite{pinho2009robust}      & 32.1±6.9        & 26.9±6.9        &                  & 96.6±4.9               \\
Bauer et al. \cite{bauer2009segmentation}      & 63.0±10.4       & 58.4±13.2       &                  & 98.6±2.1               \\
Born et al. \cite{born2009three}       & 41.7±16.2       & 34.5±13.2       &                  & 99.6±1.1               \\
Feuerstein et al. \cite{feuerstein2009adaptive} & 76.5±13.3       & 73.3±13.4       &                  & 84.4±9.5               \\
Inoue et al. \cite{inoue2013robust}      & 79.6±13.5       & 79.9±12.1       &                  & 88.1±13.2              \\
Xu et al. \cite{xu2015hybrid}         & 51.7±10.8       & 44.5±9.4        &                  & 99.2±1.6               \\
Yun et al. \cite{yun2019improvement}        & 65.7±13.1       & 60.1±11.9       &                  & 95.4±3.7               \\
Wingsnet \cite{zheng2021alleviating}          & 80.5±12.5       & 79.0±11.1       &                  & 94.2±4.3               \\
AirwayNet \cite{qin2019airwaynet}         & 76.7±11.5       & 72.7±11.6       & 3.65±2.86        &                        \\
Proposed Method   & 81.4±12.2       & 79.6±11.0       & 8.27±5.09        &                        \\ \hline
\end{tabular}
\label{table3}
\end{table}
Table \ref{table3} compared the proposed method with other SOTA algorithms on EXACT’09 dataset. The challenge organizer provided the results of our proposed method. Only BD and TD score were utilized for comparative evaluation between the first nine methods and our proposed method. And the FPR metric only compared between the AirwayNet \cite{qin2019airwaynet} method and ours. As shown in Table 5, among the results of these methods, our proposed method achieved the best BD (81.4\%) and comparative TD (79.6\%). The Wingsnet \cite{zheng2021alleviating} and Inoue et al. \cite{inoue2013robust} achieve comparative results with 80.5\% BD, 79\% TD, and 79.6\% BD, 79.9\% TD, respectively. 
\begin{figure}[H] 
    \vspace{-1em}
    \centering 
    \includegraphics[width=0.9\textwidth]{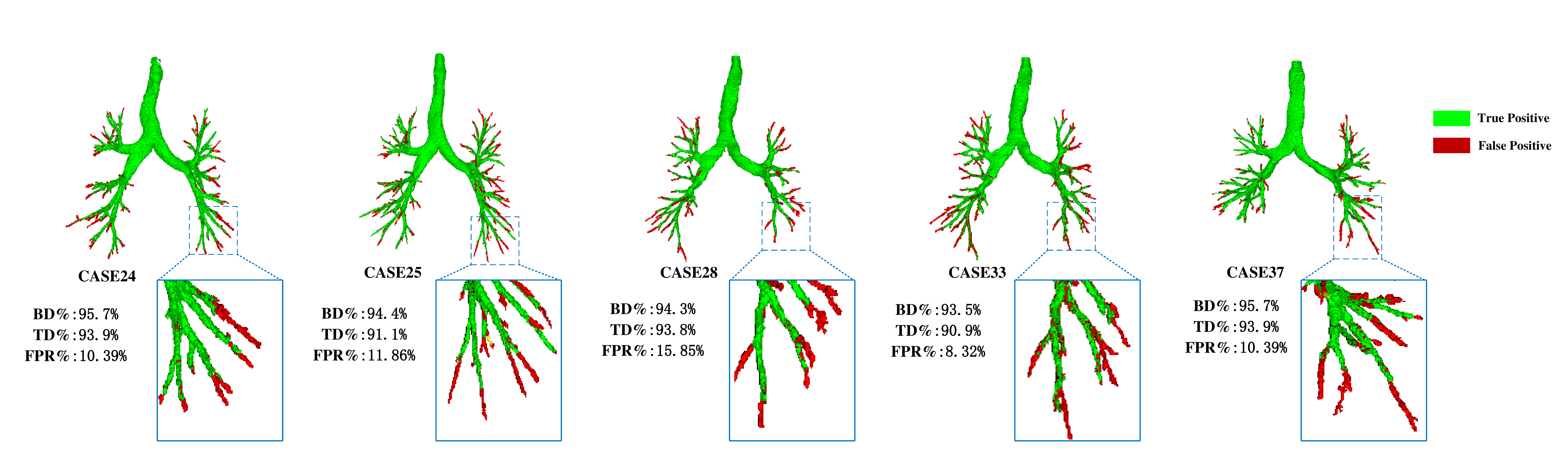} 
    \caption{Visualization of segmentation results on EXACT’09-test. The true positives (TP) and false positives (FP) are colored in green and red, respectively. The blue boxes indicate the local details magnified for better comparison.} 
    \label{fig8} 
\end{figure}
Fig. \ref{fig8} demonstrates five cases with relatively higher BD, TD, and FPR in our segmentation results. The highest FPR is observed in CASE 25, which can be detected more airway segments belong to unannotated distal bronchi without significant clumping leakage.
\subsection{Performance comparison on public BAS dataset}
For BAS dataset, the configuration of the dataset is conducted as in [35]. The 90 scans were split into 50 cases in the training set, 20 cases in the validation set, and 20 cases in the test set, respectively.
\begin{table}[H]
\caption{Performance comparison in the BAS test dataset.}
\centering
\begin{tabular}{llll}
\hline
\textit{Dataset} & \textit{TD (\%)} & \textit{BD (\%)} & \textit{Precision (\%)} \\ \hline
Wingsnet \cite{zheng2021alleviating}        & 92.5±4.5         & 88.7±7.9         & 91.4±3.3                \\
Juarez et al \cite{garcia2018automatic}    & 84.1±8.6         & 82.1±12.4        & 91.4±2.5                \\
Jin et al. \cite{jin20173d}      & 85.4±10.4        & 83.1±11.5        & 93.9±1.9                \\
Wang et al. \cite{wang2019tubular}      & 86.3±8.5         & 83.5±11.2        & 93.4±2:1                \\
Juarez et al. \cite{garcia2019joint}    & 68.0±21.1        & 60.5±23.9        & 96.4±1.8                \\
AirwayNet \cite{qin2019airwaynet}       & 83.6±10.4        & 81.4±13.8        & 95.8±1.8                \\
Proposed Method  & 94.93±3.36       & 92.4±6.01        & 86.88±4.08              \\ \hline
\end{tabular}
\label{table4}
\end{table}
Table \ref{table4} compares the proposed method with other SOTA methods on the BAS dataset. The proposed method achieves a higher TD (95.36\%) and BD (93.14\%), but a lower precision (84.15\%). WingsNet \cite{zheng2021alleviating} with group supervision for inter-class imbalance and a General Union loss for intra-class imbalance achieve the comparative result (TD: 92.5\%, BD: 88.7\%, and Precision: 91.4\%). Juarez et al. designed a modified U-Net with graph neural network architecture for better performance  with an 84.1\% TD, 82.1\% BD, and 91.4\% precision compared with their previous work, which introduced the CNN-based method with weighted binary cross entropy loss  with a 68.0\% TD, 60.54\% BD, and 96.4\% precision \cite{garcia2018automatic}. Jin et al. \cite{jin20173d} proposed the graph-based refinement for airway segmentation to work for the TD, BD, and Precision scores with 85.4\%, 83.1\%, and 93.9\%. Wang et al. \cite{wang2019tubular} train a spatial fully connected network using radial distance loss and achieves a relatively better TD (86.3\%). Qin et al. \cite{qin2019airwaynet} developed AirwayNet which predicts the connectivity of airways voxels to achieve a relative lower TD (83.6\%) and BD (81.4\%). Compared our proposed method with these works, an increment of TD between 2.68\% and 27.36\% and an increased BD between 4.44\% and 22.64\% were observed.
\begin{figure}[H] 
    \vspace{-1em}
    \centering 
    \includegraphics[width=0.9\textwidth]{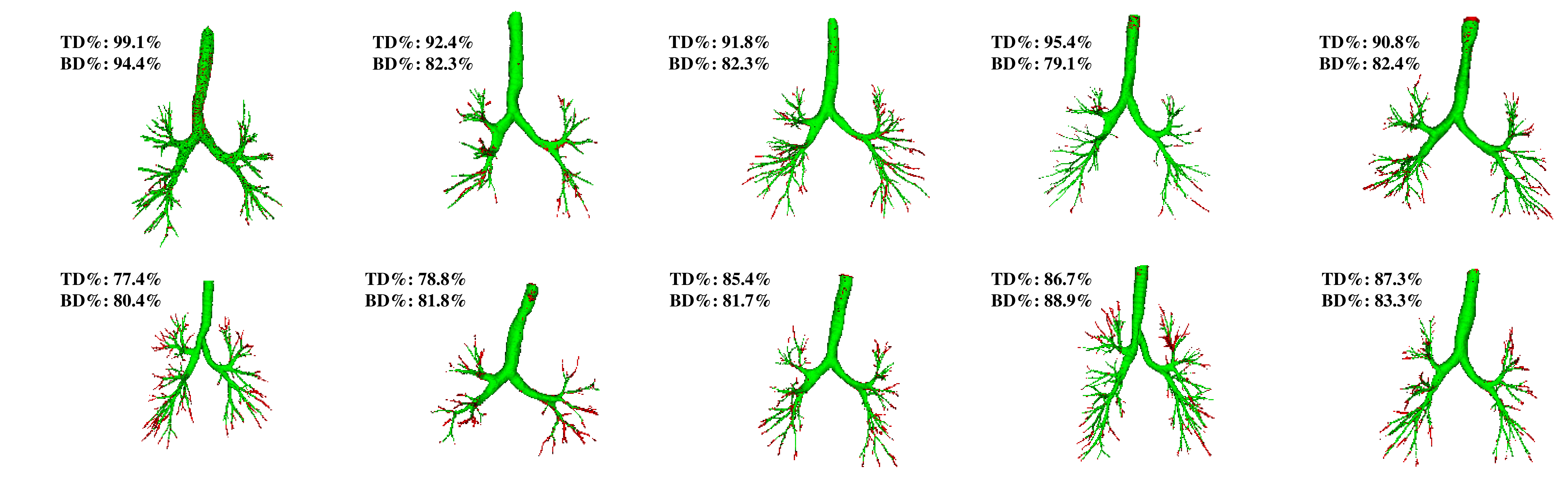} 
    \caption{Visualization of segmentation results on BAS-test. The true positives (TP) and false positives (FP) are colored in green and red, respectively.} 
    \label{fig9} 
\end{figure}
Fig. \ref{fig9} visualizes the segmentation results of the proposed method on the BAS dataset. More airway segments can be detected as unannotated distal bronchi without significant clumping leakage.
\subsection{Performance comparison on the ATM 22 dataset}

\begin{figure}[H] 
    \vspace{-1em}
    \centering 
    \includegraphics[width=0.9\textwidth]{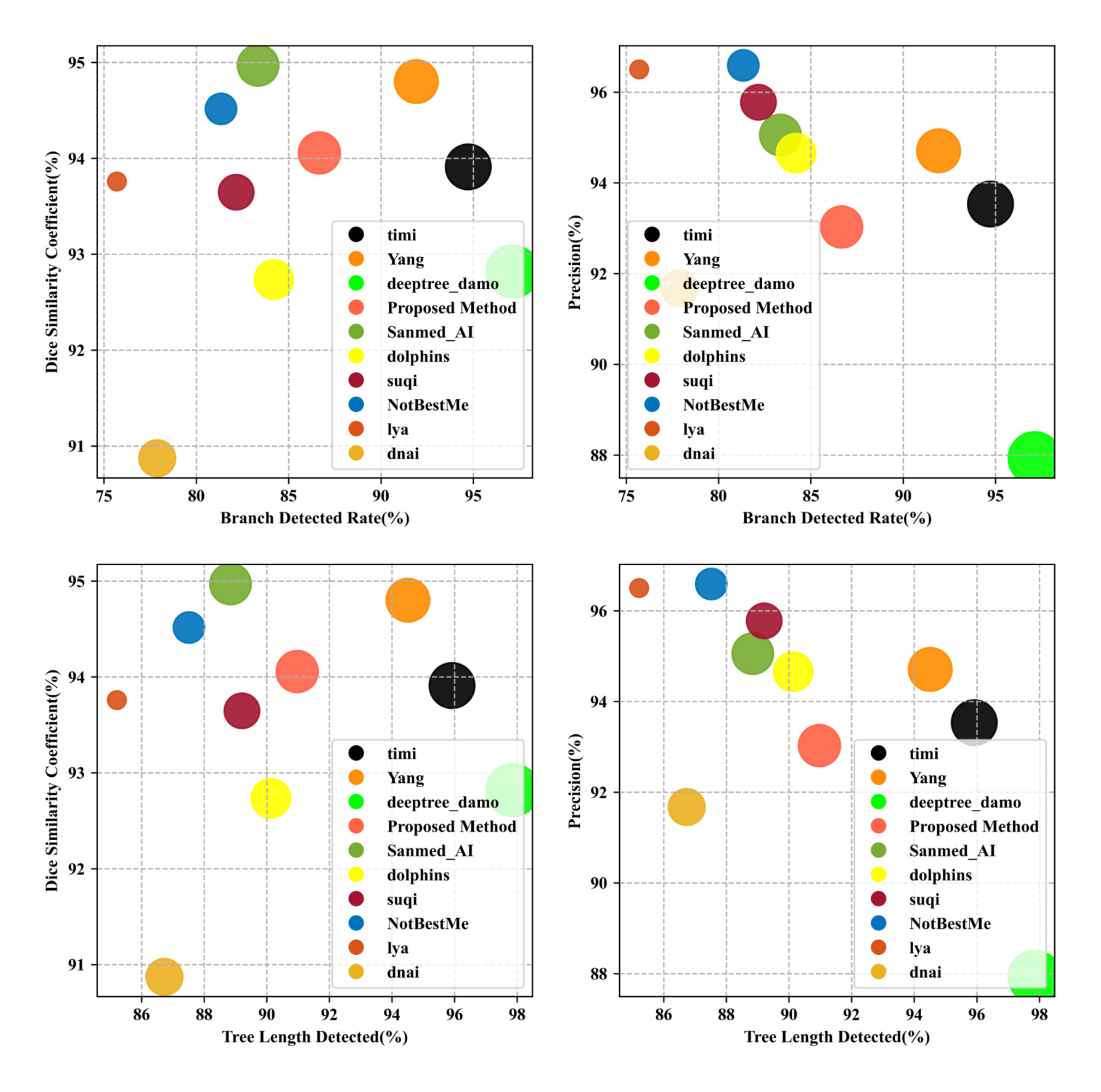} 
    \caption{Segmentation results of BD, TD, DSC, and Precision on ATM test dataset.} 
    \label{fig10} 
\end{figure}

The ATM22 dataset is conducted as the instruction as a challenge organizer on the website \href{https://atm22.grand-challenge.org/}{(https://atm22.grand-challenge.org/}). The metric of the test result is defined as Score (Score=25\%TD+25\%BD+25\%DSC+25\%Precision). As shown in Table \ref{table5} and Fig. \ref{fig10}, four evaluation metrics for ten teams were presented. The results were also posted on the challenge website (https://atm22.grand-challenge.org/final-results/). And it is also observed that the proposed method gained the fourth grade of TD, BD, and DSC with 90.974\%, 86.670\%, and 94.056\%.
In the ATM22 challenge, the best performance was observed in the timi team with a TD of 95.919\%, a BD of 94.729\%, a DSC of 93.910, and a precision of 93.553\%. Except for the DSC, the metrics of timi team exceeds the TD of our algorithm by 5\% and BD by 8\%. Moreover, we gain better performance than the dnai team with an improvement of TD (4\%), BD (9\%), DSC (3\%), and precision (2\%).
\begin{table}[H]
\caption{Performance comparison on the ATM 22 dataset (Test Phase)}
\centering
\begin{tabular}{llllll}
\hline
\textit{Team}  & \textit{TD (\%)} & \textit{BD (\%)} & \textit{DSC} & Precision (\%) & Score  \\ \hline
timi           & 95.919           & 94.729           & 93.910       & 93.553         & 94.528 \\
Yang           & 94.512           & 91.920           & 94.800       & 94.707         & 93.985 \\
deeptree\_damo & 97.853           & 97.129           & 92.819       & 87.928         & 93.932 \\
our\_proposed  & 90.974           & 86.670           & 94.056       & 93.027         & 91.182 \\
Sanmed\_AI     & 88.843           & 83.350           & 94.969       & 95.055         & 90.554 \\
dolphins       & 90.134           & 84.201           & 92.734       & 94.656         & 90.431 \\
suqi           & 89.209           & 82.164           & 93.646       & 95.777         & 90.199 \\
NotBestMe      & 87.518           & 81.343           & 94.515       & 96.590         & 89.992 \\
lya            & 85.215           & 75.705           & 93.758       & 96.501         & 87.795 \\
dnai           & 86.733           & 77.888           & 90.871       & 91.674         & 86.792 \\ \hline
\end{tabular}
\label{table5}
\end{table}
\subsection{Application of proposed methods on COPD and COVID-19 datasets}
\begin{figure}[H] 
    \vspace{-1em}
    \centering 
    \includegraphics[width=0.9\textwidth]{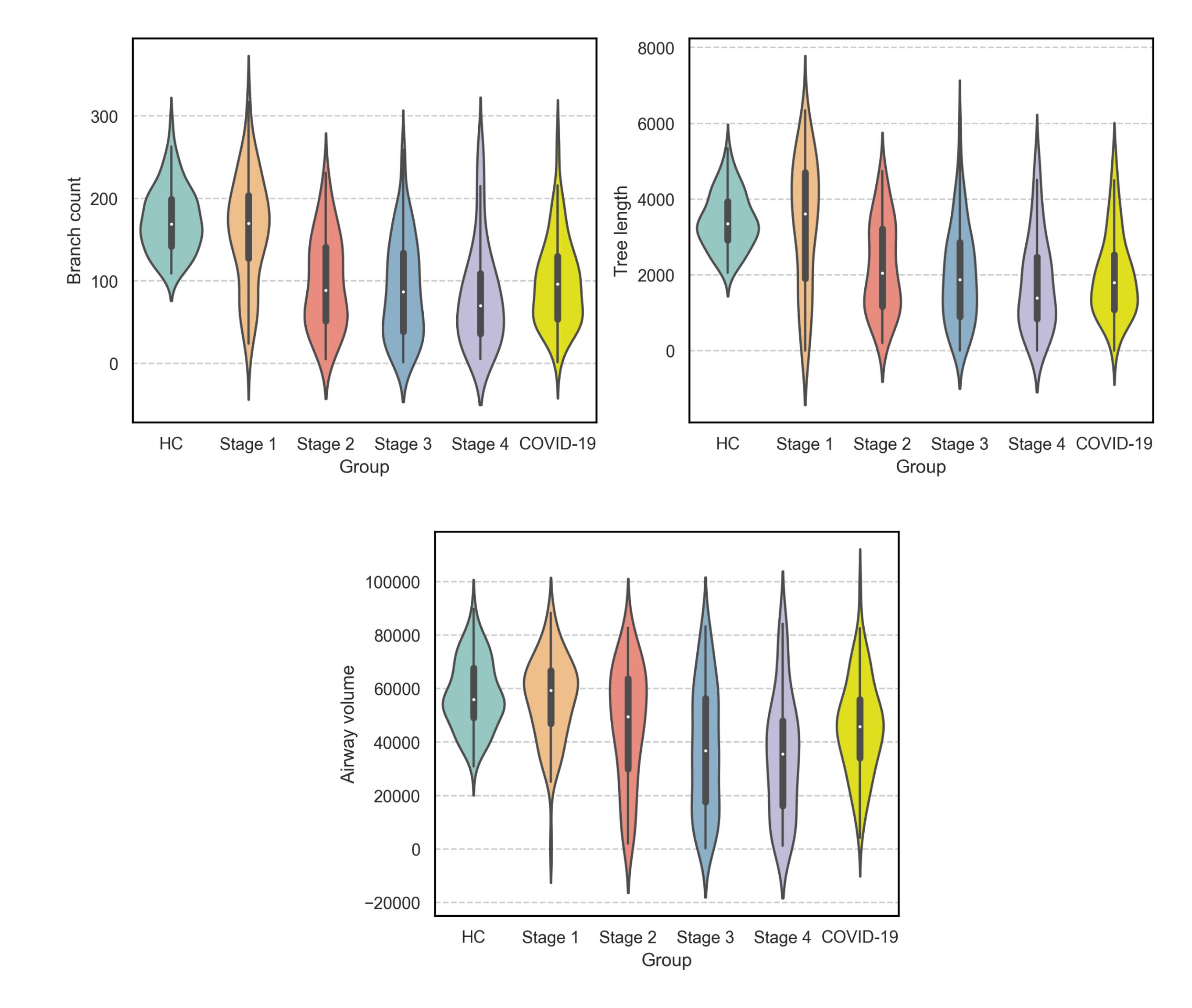} 
    \caption{Comparison performance of HC with COPD and COVID-19 in branch count, tree length, and airway volume metric.} 
    \label{fig11} 
\end{figure}
Table \ref{table6} and Fig. \ref{fig11} presents the quantitative performance of branch count, tree length, and airway volume on a different stage of COPD (row 1-4) and COVID-19 (row 5) datasets. Because the labels of these datasets are not acquired, branch count, tree length (mm), and airway volume (mm3) as they were more reference-independent are employed for performance comparison of our method on different stages of COPD and the COVID-19 dataset. Results showed that the detected branch count and tree length are on the decrease and the airway volume decreased as more severe COPD. Moreover, our method achieved the performance of Branch count with 98.9±53.98, Tree length (mm) at 1974.56±1120.59, and Airway volume (mm3) at 45594.49±17949.17.
\begin{table}[H]
\caption{Performance of branch count, tree length, and airway volume on COPD and COVID-19 dataset.}
\centering
\begin{tabular}{llll}
\hline
\textit{Group} & \textit{Branch count} & \textit{Tree length(mm)} & \textit{Airway volume(mm3)} \\ \hline
HC             & 174.04±39.56          & 3427.18±747.41           & 57778.05±13047.11           \\
COPD Stage \uppercase\expandafter{\romannumeral1}   & 161.29±67.06          & 3373.58±1718.99          & 56310.70±15799.92           \\
COPD Stage \uppercase\expandafter{\romannumeral2}   & 97.88±57.19           & 2182.75±1211.817         & 46537.17±21863.75           \\
COPD Stage \uppercase\expandafter{\romannumeral3}   & 90.87±61.83           & 1994.41±1299.19          & 37819.44±23697.17           \\
COPD Stage \uppercase\expandafter{\romannumeral4}   & 82.59±64.69           & 1796.15±1279.89          & 34943.82±22902.42           \\
COVID-19       & 98.19±53.98           & 1974.56±1120.59          & 45594.49±17949.17           \\ \hline
\end{tabular}
\label{table6}
\end{table}

\begin{figure}[H] 
    \vspace{-1em}
    \centering 
    \includegraphics[width=0.6\textwidth]{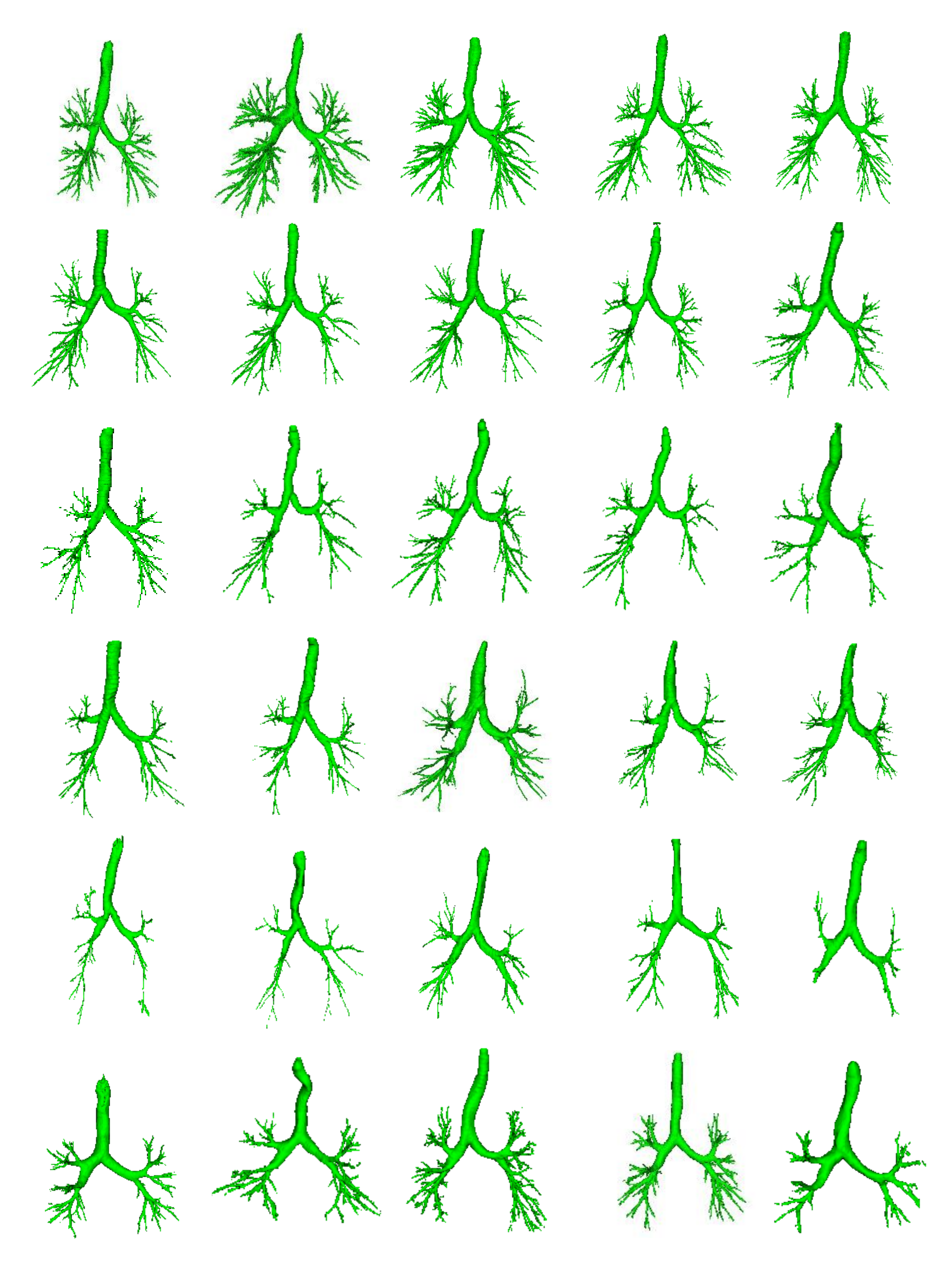} 
    \caption{Exemplary segmentation results in applications. From the first row: HC, COPD GOLD stage I, COPD stage II, COPD GOLD stage III, COPD GOLD stage IV, and COVID-19.} 
    \label{fig12} 
\end{figure}

Fig. \ref{fig12} presents the visualization of segmentation results in the COPD and COVID-19 dataset. With the severity of different stages of COPD, it is observed that the airway tree exhibits morphological defects, including less bronchial bronchus, a rougher surface, and abruptly significant stenosis. Emphysema or inflammation in COPD sufferers may be the cause of it. In the COVID-19 dataset, we also visualize the many bronchi and good completeness in the results, which can determine that our proposed two-stage and CoT block boost the segmentation performance of pulmonary airways.
\section{Discussion}
Accurate segmentation of airway from CT images from constrained labels is challenging work. This study uses a two-stage method with the same 3D contextual transformer-based U-Net subnetwork to perform initial and refined airway segmentation. The performance on in-house and multiple public datasets prove that our proposed method extracted much more branches and lengths of the tree and accomplished state-of-the-art airway segmentation performance. 
\begin{figure}[H] 
    \vspace{-1em}
    \centering 
    \includegraphics[width=0.7\textwidth]{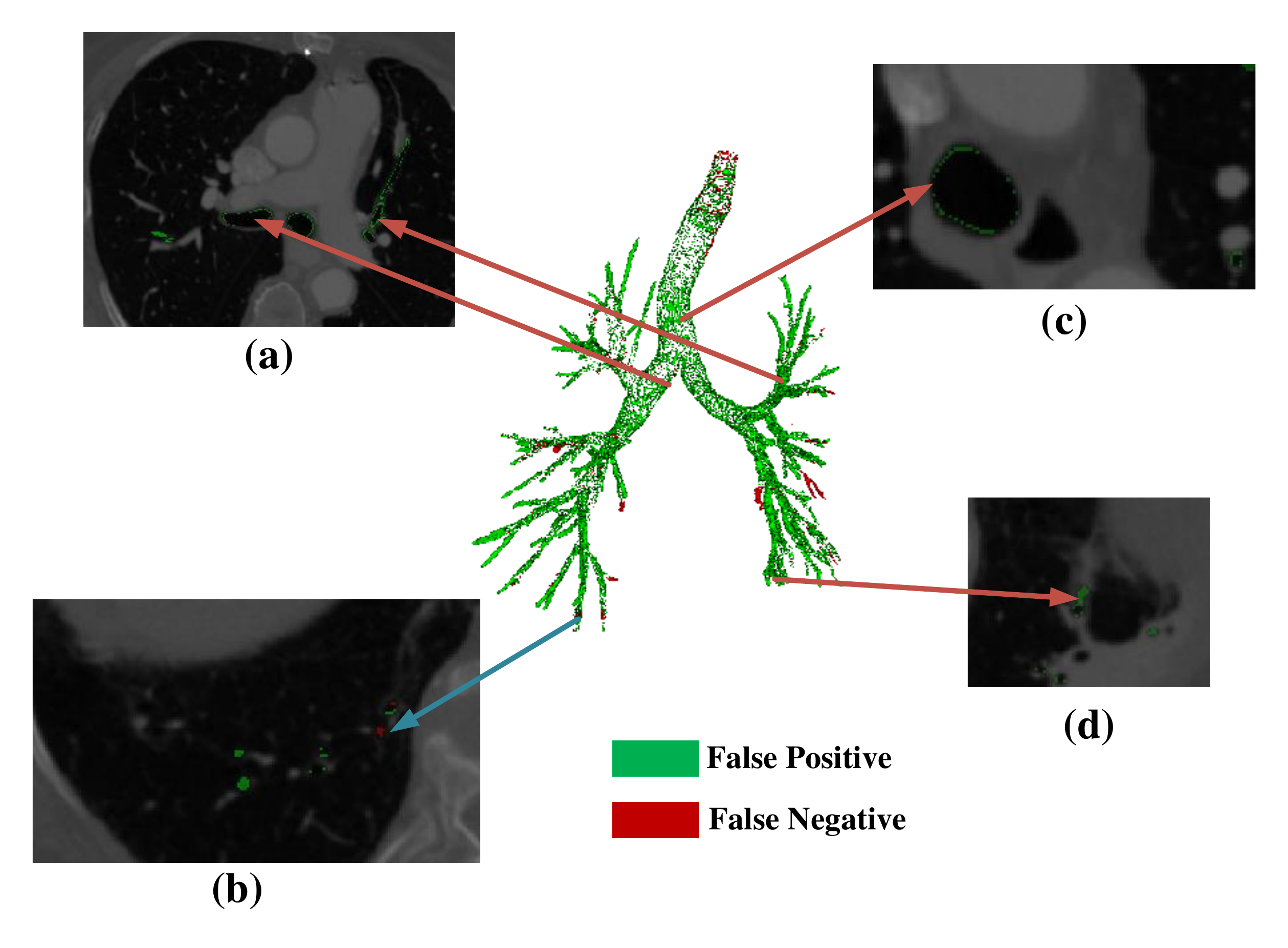} 
    \caption{False positives and false negatives in the segmentation results. Five representative regions are selected, in which red arrows demonstrate the false positives corresponding with the original CT images and blue arrows demonstrate the false negatives corresponding with the original CT images.} 
    \label{fig13} 
\end{figure}
The main improvement of the proposed method is the tree branch, and tree length detected, along with the decline in precision. Compared with AirwayNet \cite{qin2019airwaynet}, our proposed method improves the tree branch and tree length detected by about 5\% and 9\% and decreases the DSC by 2\% on ISICDM 2021 dataset. Moreover, our proposed method improves the tree branch and tree length detected by about 5\% and 7\% on the EXACT’09 dataset and improves the tree branch detected by about 11\% and tree length detected by about 11\%, but decreases the precision by about 9\% on BAS dataset. AirwayNet was proposed to predict airways voxels' connectivity, which improves the precision, but the tree branch and tree length detected are still limited. The refined network in our method is an excellent solution for this issue. Compared with WingsNet \cite{zheng2021alleviating}, the tree length detected improved by about 5\%, 1\%, and 3\%, and the tree branch detected improved by about 3\%, 1\%, and 4\% on ISICDM 2021, EXACT’09, and BAS datasets. Compared with the above two well-liked algorithms, our proposed method achieved success in the branch and tree length detected. However, this will inevitably lead to false positives. An example is shown in Fig. \ref{fig13} from the BAS dataset to explain the phenomenon. As described in Fig. \ref{fig13} (a) and (c), most of the FPs appear in the margins of the trachea and main airway. Accurately manual annotation of the airway is a difficult and complex work, and keeping the connectivity is the crucial step. Therefore, some unannotated pixels will happen. Also, some annotations of the airway may be beyond the reach of radiologists, as shown in Fig. \ref{fig13}(d), Our proposed refined network could segment these high-generation airways. However, the refined network may cause several over-segmentations (Fig. \ref{fig13}(c)).

The proposed method was also applied in the COPD and COVID-19 datasets for airway segmentation. The results described in Table 6 can also present the quantitative airway change in a different stage of COPD patients. In COPD patients, increased thickness of airway walls, airflow limitation, and luminal obstruction by muco-inflammatory exudates can independently contribute to the airflow limitation and could be the main factors of change of airway structural abnormalities \cite{crisafulli2017prevalence}. This may be the main reason for the decreased detection of branch count, tree length, and airway volume. In COVID-19 patients, ground glass opacities, consolidation, and severity of opacifications \cite{xiong2020clinical} can be observed in CT images, which increases the difficulties of airway segmentation. Table 6, and Figure 11 and 12 demonstrate that our proposed method achieved promising performance in airway segmentation. The above application further proved that the proposed algorithms are robust and widely applicable.

The study provided an accurate and robust method with two-stage CoT-based convolutional neural networks for airway segmentation. However, the dataset with label datasets of specific diseases, such as COVID-19 and COPD, is not included in our study. In the next step, these situations should be considered in our future work. Second, the U-Net with 3D CoT block is only used for airway segmentation from CT images. More applications from volumetric data, such as vessel segmentatrion, artery-vein segmentation, and lung cancer lesions should be taken into account. Moreover, labeling airway from CT images is difficult, complicated, and time-consuming, the semi-supervised learning and self-semi-supervised learning will be explored in future work.
\section{Conclusion}
A novel two-stage 3D contextual transformer-based U-Net is proposed for airway segmentation from CT images. The method consists of two stages performing initial and refined airway segmentation, in which the same subnetwork with different airway masks as input is employed. Contextual transformer block and a two-stage strategy are performed to effectively improve the quality of airway segmentation. Extensive experiments demonstrate that our proposed method extracted much more branches and lengths of the tree while accomplishing state-of-the-art airway segmentation performance. The proposed method could contribute to planning navigation bronchoscopy and quantitative assessment of airway-related chronic obstructive pulmonary disease.

\bibliographystyle{unsrt}
\bibliography{references}  

\begin{thebibliography}{10}

\bibitem{benn2021robotic}
Bryan~S Benn, Arthur~O Romero, Mendy Lum, and Ganesh Krishna.
\newblock Robotic-assisted navigation bronchoscopy as a paradigm shift in
  peripheral lung access.
\newblock {\em Lung}, 199(2):177--186, 2021.

\bibitem{tanabe2021central}
Naoya Tanabe, Kaoruko Shimizu, Kunihiko Terada, Susumu Sato, Masaru Suzuki,
  Hiroshi Shima, Akira Oguma, Tsuyoshi Oguma, Satoshi Konno, Masaharu
  Nishimura, et~al.
\newblock Central airway and peripheral lung structures in airway
  disease-dominant copd.
\newblock {\em ERJ open research}, 7(1), 2021.

\bibitem{lu2021necroptosis}
Zhe Lu, Hannelore~P Van~Eeckhoutte, Gang Liu, Prema~M Nair, Bernadette Jones,
  Caitlin~M Gillis, B~Christina Nalkurthi, Fien Verhamme, Tamariche
  Buyle-Huybrecht, Peter Vandenabeele, et~al.
\newblock Necroptosis signaling promotes inflammation, airway remodeling, and
  emphysema in chronic obstructive pulmonary disease.
\newblock {\em American journal of respiratory and critical care medicine},
  204(6):667--681, 2021.

\bibitem{ishiwata2020bronchoscopic}
Tsukasa Ishiwata, Alexander Gregor, Terunaga Inage, and Kazuhiro Yasufuku.
\newblock Bronchoscopic navigation and tissue diagnosis.
\newblock {\em General thoracic and cardiovascular surgery}, 68(7):672--678,
  2020.

\bibitem{kemp2020navigation}
Samuel~Victor Kemp.
\newblock Navigation bronchoscopy.
\newblock {\em Respiration}, 99(4):277--286, 2020.

\bibitem{asano2014virtual}
Fumihiro Asano, Ralf Eberhardt, and Felix~JF Herth.
\newblock Virtual bronchoscopic navigation for peripheral pulmonary lesions.
\newblock {\em Respiration}, 88(5):430--440, 2014.

\bibitem{edell2010navigational}
Eric Edell and Diane Krier-Morrow.
\newblock Navigational bronchoscopy: Overview of technology and practical
  considerations—new current procedural terminology codes effective 2010.
\newblock {\em Chest}, 137(2):450--454, 2010.

\bibitem{shen2019context}
Mali Shen, Yun Gu, Ning Liu, and Guang-Zhong Yang.
\newblock Context-aware depth and pose estimation for bronchoscopic navigation.
\newblock {\em IEEE Robotics and Automation Letters}, 4(2):732--739, 2019.

\bibitem{mehta2018evolutional}
Atul~C Mehta, Kristin~L Hood, Yehuda Schwarz, and Stephen~B Solomon.
\newblock The evolutional history of electromagnetic navigation bronchoscopy:
  state of the art.
\newblock {\em Chest}, 154(4):935--947, 2018.

\bibitem{higgins2015multimodal}
William~E Higgins, Ronnarit Cheirsilp, Xiaonan Zang, and Patrick Byrnes.
\newblock Multimodal system for the planning and guidance of bronchoscopy.
\newblock In {\em Medical Imaging 2015: Image-Guided Procedures, Robotic
  Interventions, and Modeling}, volume 9415, pages 43--51. SPIE, 2015.

\bibitem{halpin2021global}
David~MG Halpin, Gerard~J Criner, Alberto Papi, Dave Singh, Antonio Anzueto,
  Fernando~J Martinez, Alvar~A Agusti, and Claus~F Vogelmeier.
\newblock Global initiative for the diagnosis, management, and prevention of
  chronic obstructive lung disease. the 2020 gold science committee report on
  covid-19 and chronic obstructive pulmonary disease.
\newblock {\em American journal of respiratory and critical care medicine},
  203(1):24--36, 2021.

\bibitem{hirota2013mechanisms}
Nobuaki Hirota and James~G Martin.
\newblock Mechanisms of airway remodeling.
\newblock {\em Chest}, 144(3):1026--1032, 2013.

\bibitem{ding2016measuring}
Ming Ding, Yu~Chen, Wei-Jie Guan, Chang-Hao Zhong, Mei Jiang, Wei-Zhan Luo,
  Xiao-Bo Chen, Chun-Li Tang, Yan Tang, Qi-Ming Jian, et~al.
\newblock Measuring airway remodeling in patients with different copd staging
  using endobronchial optical coherence tomography.
\newblock {\em Chest}, 150(6):1281--1290, 2016.

\bibitem{goddard1982computed}
Paul~R Goddard, EM~Nicholson, G~Laszlo, and I~Watt.
\newblock Computed tomography in pulmonary emphysema.
\newblock {\em Clinical radiology}, 33(4):379--387, 1982.

\bibitem{mets2013diagnosis}
Onno~M Mets, Michael Schmidt, Constantinus~F Buckens, Martijn~J Gondrie, Ivana
  Isgum, Matthijs Oudkerk, Rozemarijn Vliegenthart, Harry~J de~Koning,
  Carlijn~M van~der Aalst, Mathias Prokop, et~al.
\newblock Diagnosis of chronic obstructive pulmonary disease in lung cancer
  screening computed tomography scans: independent contribution of emphysema,
  air trapping and bronchial wall thickening.
\newblock {\em Respiratory research}, 14(1):1--8, 2013.

\bibitem{sasaki2014ratios}
Tomoaki Sasaki, Koji Takahashi, Nobuhisa Takada, and Yoshinobu Ohsaki.
\newblock Ratios of peripheral-to-central airway lumen area and percentage wall
  area as predictors of severity of chronic obstructive pulmonary disease.
\newblock {\em American Journal of Roentgenology}, 203(1):78--84, 2014.

\bibitem{lutey2013accurate}
Barbara~A Lutey, Susan~H Conradi, Jeffrey~J Atkinson, Jie Zheng, Kenneth~B
  Schechtman, Robert~M Senior, and David~S Gierada.
\newblock Accurate measurement of small airways on low-dose thoracic ct scans
  in smokers.
\newblock {\em Chest}, 143(5):1321--1329, 2013.

\bibitem{litjens2017survey}
Geert Litjens, Thijs Kooi, Babak~Ehteshami Bejnordi, Arnaud Arindra~Adiyoso
  Setio, Francesco Ciompi, Mohsen Ghafoorian, Jeroen~Awm Van Der~Laak, Bram
  Van~Ginneken, and Clara~I S{\'a}nchez.
\newblock A survey on deep learning in medical image analysis.
\newblock {\em Medical image analysis}, 42:60--88, 2017.

\bibitem{lecun2015deep}
Yann LeCun, Yoshua Bengio, Geoffrey Hinton, et~al.
\newblock Deep learning. nature, 521 (7553), 436-444.
\newblock {\em Google Scholar Google Scholar Cross Ref Cross Ref}, 2015.

\bibitem{luo2022word}
Xiangde Luo, Wenjun Liao, Jianghong Xiao, Jieneng Chen, Tao Song, Xiaofan
  Zhang, Kang Li, Dimitris~N Metaxas, Guotai Wang, and Shaoting Zhang.
\newblock Word: A large scale dataset, benchmark and clinical applicable study
  for abdominal organ segmentation from ct image.
\newblock {\em Medical Image Analysis}, 82:102642, 2022.

\bibitem{ma2022fast}
Jun Ma, Yao Zhang, Song Gu, Xingle An, Zhihe Wang, Cheng Ge, Congcong Wang, Fan
  Zhang, Yu~Wang, Yinan Xu, et~al.
\newblock Fast and low-gpu-memory abdomen ct organ segmentation: The flare
  challenge.
\newblock {\em Medical Image Analysis}, 82:102616, 2022.

\bibitem{chen2021diverse}
Xu~Chen, Chunfeng Lian, Li~Wang, Hannah Deng, Tianshu Kuang, Steve~H Fung,
  Jaime Gateno, Dinggang Shen, James~J Xia, and Pew-Thian Yap.
\newblock Diverse data augmentation for learning image segmentation with
  cross-modality annotations.
\newblock {\em Medical Image Analysis}, 71:102060, 2021.

\bibitem{momin2022mutual}
Shadab Momin, Yang Lei, Neal~S McCall, Jiahan Zhang, Justin Roper, Joseph
  Harms, Sibo Tian, Michael~S Lloyd, Tian Liu, Jeffrey~D Bradley, et~al.
\newblock Mutual enhancing learning-based automatic segmentation of ct cardiac
  substructure.
\newblock {\em Physics in Medicine \& Biology}, 67(10):105008, 2022.

\bibitem{feng2022bla}
Ruiqi Feng, Li~Zhuo, Xiaoguang Li, Hongxia Yin, and Zhenchang Wang.
\newblock Bla-net: Boundary learning assisted network for skin lesion
  segmentation.
\newblock {\em Computer Methods and Programs in Biomedicine}, 226:107190, 2022.

\bibitem{wu2021elnet}
Zhan Wu, Rongjun Ge, Minli Wen, Gaoshuang Liu, Yang Chen, Pinzheng Zhang,
  Xiaopu He, Jie Hua, Limin Luo, and Shuo Li.
\newblock Elnet: Automatic classification and segmentation for esophageal
  lesions using convolutional neural network.
\newblock {\em Medical Image Analysis}, 67:101838, 2021.

\bibitem{chen2021effective}
Cheng Chen, Kangneng Zhou, Muxi Zha, Xiangyan Qu, Xiaoyu Guo, Hongyu Chen,
  Zhiliang Wang, and Ruoxiu Xiao.
\newblock An effective deep neural network for lung lesions segmentation from
  covid-19 ct images.
\newblock {\em IEEE Transactions on Industrial Informatics}, 17(9):6528--6538,
  2021.

\bibitem{dosovitskiy2020image}
Alexey Dosovitskiy, Lucas Beyer, Alexander Kolesnikov, Dirk Weissenborn,
  Xiaohua Zhai, Thomas Unterthiner, Mostafa Dehghani, Matthias Minderer, Georg
  Heigold, Sylvain Gelly, et~al.
\newblock An image is worth 16x16 words: Transformers for image recognition at
  scale.
\newblock {\em arXiv preprint arXiv:2010.11929}, 2020.

\bibitem{devlin2018bert}
Jacob Devlin, Ming-Wei Chang, Kenton Lee, and Kristina Toutanova.
\newblock Bert: Pre-training of deep bidirectional transformers for language
  understanding.
\newblock {\em arXiv preprint arXiv:1810.04805}, 2018.

\bibitem{vaswani2017attention}
Ashish Vaswani, Noam Shazeer, Niki Parmar, Jakob Uszkoreit, Llion Jones,
  Aidan~N Gomez, {\L}ukasz Kaiser, and Illia Polosukhin.
\newblock Attention is all you need.
\newblock {\em Advances in neural information processing systems}, 30, 2017.

\bibitem{valanarasu2021medical}
Jeya Maria~Jose Valanarasu, Poojan Oza, Ilker Hacihaliloglu, and Vishal~M
  Patel.
\newblock Medical transformer: Gated axial-attention for medical image
  segmentation.
\newblock In {\em International Conference on Medical Image Computing and
  Computer-Assisted Intervention}, pages 36--46. Springer, 2021.

\bibitem{wu2021vision}
Yanan Wu, Shouliang Qi, Yu~Sun, Shuyue Xia, Yudong Yao, and Wei Qian.
\newblock A vision transformer for emphysema classification using ct images.
\newblock {\em Physics in Medicine \& Biology}, 66(24):245016, 2021.

\bibitem{zhao2022cot}
Shuiqing Zhao, Yanan Wu, Mengmeng Tong, Yudong Yao, Wei Qian, and Shouliang Qi.
\newblock Cot-xnet: Contextual transformer with xception network for diabetic
  retinopathy grading.
\newblock {\em Physics in Medicine \& Biology}, 2022.

\bibitem{raghu2021vision}
Maithra Raghu, Thomas Unterthiner, Simon Kornblith, Chiyuan Zhang, and Alexey
  Dosovitskiy.
\newblock Do vision transformers see like convolutional neural networks?
\newblock {\em Advances in Neural Information Processing Systems},
  34:12116--12128, 2021.

\bibitem{li2022contextual}
Yehao Li, Ting Yao, Yingwei Pan, and Tao Mei.
\newblock Contextual transformer networks for visual recognition.
\newblock {\em IEEE Transactions on Pattern Analysis and Machine Intelligence},
  2022.

\bibitem{zheng2021alleviating}
Hao Zheng, Yulei Qin, Yun Gu, Fangfang Xie, Jie Yang, Jiayuan Sun, and
  Guang-Zhong Yang.
\newblock Alleviating class-wise gradient imbalance for pulmonary airway
  segmentation.
\newblock {\em IEEE Transactions on Medical Imaging}, 40(9):2452--2462, 2021.

\bibitem{wang2022naviairway}
Andong Wang, Terence Chi~Chun Tam, Ho~Ming Poon, Kun-Chang Yu, and Wei-Ning
  Lee.
\newblock Naviairway: a bronchiole-sensitive deep learning-based airway
  segmentation pipeline for planning of navigation bronchoscopy.
\newblock {\em arXiv preprint arXiv:2203.04294}, 2022.

\bibitem{fabijanska2009two}
Anna Fabija{\'n}ska.
\newblock Two-pass region growing algorithm for segmenting airway tree from
  mdct chest scans.
\newblock {\em Computerized Medical Imaging and Graphics}, 33(7):537--546,
  2009.

\bibitem{pinho2009robust}
R{\^o}mulo Pinho, Sten Luyckx, and Jan Sijbers.
\newblock Robust region growing based intrathoracic airway tree segmentation.
\newblock In {\em Proc. of Second International Workshop on Pulmonary Image
  Analysis}, pages 261--271, 2009.

\bibitem{shi2006upper}
Hongjian Shi, William~C Scarfe, and Allan~G Farman.
\newblock Upper airway segmentation and dimensions estimation from cone-beam ct
  image datasets.
\newblock {\em International Journal of Computer Assisted Radiology and
  Surgery}, 1(3):177--186, 2006.

\bibitem{aykac2003segmentation}
Deniz Aykac, Eric~A Hoffman, Geoffrey McLennan, and Joseph~M Reinhardt.
\newblock Segmentation and analysis of the human airway tree from
  three-dimensional x-ray ct images.
\newblock {\em IEEE transactions on medical imaging}, 22(8):940--950, 2003.

\bibitem{born2009three}
Silvia Born, Dirk Iwamaru, Matthias Pfeifle, and Dirk Bartz.
\newblock Three-step segmentation of the lower airways with advanced
  leakage-control.
\newblock In {\em Proc. of Second International Workshop on Pulmonary Image
  Analysis}, pages 239--249. Citeseer, 2009.

\bibitem{bauer2009segmentation}
Christian Bauer, Horst Bischof, and Reinhard Beichel.
\newblock Segmentation of airways based on gradient vector flow.
\newblock In {\em International workshop on pulmonary image analysis, Medical
  image computing and computer assisted intervention}, pages 191--201, 2009.

\bibitem{tschirren2005segmentation}
Juerg Tschirren, Eric~A Hoffman, Geoffrey McLennan, and Milan Sonka.
\newblock Segmentation and quantitative analysis of intrathoracic airway trees
  from computed tomography images.
\newblock {\em Proceedings of the American Thoracic Society}, 2(6):484--487,
  2005.

\bibitem{tschirren2005intrathoracic}
Juerg Tschirren, Eric~A Hoffman, Geoffrey McLennan, and Milan Sonka.
\newblock Intrathoracic airway trees: segmentation and airway morphology
  analysis from low-dose ct scans.
\newblock {\em IEEE transactions on medical imaging}, 24(12):1529--1539, 2005.

\bibitem{kiraly2002three}
Atilla~P Kiraly, William~E Higgins, Geoffrey McLennan, Eric~A Hoffman, and
  Joseph~M Reinhardt.
\newblock Three-dimensional human airway segmentation methods for clinical
  virtual bronchoscopy.
\newblock {\em Academic radiology}, 9(10):1153--1168, 2002.

\bibitem{meng2017automatic}
Qier Meng, Takayuki Kitasaka, Yukitaka Nimura, Masahiro Oda, Junji Ueno, and
  Kensaku Mori.
\newblock Automatic segmentation of airway tree based on local intensity filter
  and machine learning technique in 3d chest ct volume.
\newblock {\em International journal of computer assisted radiology and
  surgery}, 12(2):245--261, 2017.

\bibitem{lo2012extraction}
Pechin Lo, Bram Van~Ginneken, Joseph~M Reinhardt, Tarunashree Yavarna, Pim~A
  De~Jong, Benjamin Irving, Catalin Fetita, Margarete Ortner, R{\^o}mulo Pinho,
  Jan Sijbers, et~al.
\newblock Extraction of airways from ct (exact'09).
\newblock {\em IEEE Transactions on Medical Imaging}, 31(11):2093--2107, 2012.

\bibitem{ronneberger2015u}
Olaf Ronneberger, Philipp Fischer, and Thomas Brox.
\newblock U-net: Convolutional networks for biomedical image segmentation.
\newblock In {\em International Conference on Medical image computing and
  computer-assisted intervention}, pages 234--241. Springer, 2015.

\bibitem{nadeem2020ct}
Syed~Ahmed Nadeem, Eric~A Hoffman, Jessica~C Sieren, Alejandro~P Comellas,
  Surya~P Bhatt, Igor~Z Barjaktarevic, Fereidoun Abtin, and Punam~K Saha.
\newblock A ct-based automated algorithm for airway segmentation using
  freeze-and-grow propagation and deep learning.
\newblock {\em IEEE transactions on medical imaging}, 40(1):405--418, 2020.

\bibitem{garcia2018automatic}
Antonio Garcia-Uceda~Juarez, Harm~AWM Tiddens, and M~de Bruijne.
\newblock Automatic airway segmentation in chest ct using convolutional neural
  networks.
\newblock In {\em Image analysis for moving organ, breast, and thoracic
  images}, pages 238--250. Springer, 2018.

\bibitem{jin20173d}
Dakai Jin, Ziyue Xu, Adam~P Harrison, Kevin George, and Daniel~J Mollura.
\newblock 3d convolutional neural networks with graph refinement for airway
  segmentation using incomplete data labels.
\newblock In {\em International workshop on machine learning in medical
  imaging}, pages 141--149. Springer, 2017.

\bibitem{wang2019tubular}
Chenglong Wang, Yuichiro Hayashi, Masahiro Oda, Hayato Itoh, Takayuki Kitasaka,
  Alejandro~F Frangi, and Kensaku Mori.
\newblock Tubular structure segmentation using spatial fully connected network
  with radial distance loss for 3d medical images.
\newblock In {\em International Conference on Medical Image Computing and
  Computer-Assisted Intervention}, pages 348--356. Springer, 2019.

\bibitem{garcia2019joint}
Antonio Garcia-Uceda~Juarez, Raghavendra Selvan, Zaigham Saghir, and Marleen~de
  Bruijne.
\newblock A joint 3d unet-graph neural network-based method for airway
  segmentation from chest cts.
\newblock In {\em International workshop on machine learning in medical
  imaging}, pages 583--591. Springer, 2019.

\bibitem{selvan2020graph}
Raghavendra Selvan, Thomas Kipf, Max Welling, Antonio Garcia-Uceda Juarez,
  Jesper~H Pedersen, Jens Petersen, and Marleen de~Bruijne.
\newblock Graph refinement based airway extraction using mean-field networks
  and graph neural networks.
\newblock {\em Medical Image Analysis}, 64:101751, 2020.

\bibitem{qin2021learning}
Yulei Qin, Hao Zheng, Yun Gu, Xiaolin Huang, Jie Yang, Lihui Wang, Feng Yao,
  Yue-Min Zhu, and Guang-Zhong Yang.
\newblock Learning tubule-sensitive cnns for pulmonary airway and artery-vein
  segmentation in ct.
\newblock {\em IEEE Transactions on Medical Imaging}, 40(6):1603--1617, 2021.

\bibitem{nan2022fuzzy}
Yang Nan, Javier Del~Ser, Zeyu Tang, Peng Tang, Xiaodan Xing, Yingying Fang,
  Francisco Herrera, Witold Pedrycz, Simon Walsh, and Guang Yang.
\newblock Fuzzy attention neural network to tackle discontinuity in airway
  segmentation.
\newblock {\em arXiv preprint arXiv:2209.02048}, 2022.

\bibitem{chen2021transunet}
Jieneng Chen, Yongyi Lu, Qihang Yu, Xiangde Luo, Ehsan Adeli, Yan Wang, Le~Lu,
  Alan~L Yuille, and Yuyin Zhou.
\newblock Transunet: Transformers make strong encoders for medical image
  segmentation.
\newblock {\em arXiv preprint arXiv:2102.04306}, 2021.

\bibitem{liu2021swin}
Ze~Liu, Yutong Lin, Yue Cao, Han Hu, Yixuan Wei, Zheng Zhang, Stephen Lin, and
  Baining Guo.
\newblock Swin transformer: Hierarchical vision transformer using shifted
  windows.
\newblock In {\em Proceedings of the IEEE/CVF International Conference on
  Computer Vision}, pages 10012--10022, 2021.

\bibitem{xie2021segformer}
Enze Xie, Wenhai Wang, Zhiding Yu, Anima Anandkumar, Jose~M Alvarez, and Ping
  Luo.
\newblock Segformer: Simple and efficient design for semantic segmentation with
  transformers.
\newblock {\em Advances in Neural Information Processing Systems},
  34:12077--12090, 2021.

\bibitem{zheng2021rethinking}
Sixiao Zheng, Jiachen Lu, Hengshuang Zhao, Xiatian Zhu, Zekun Luo, Yabiao Wang,
  Yanwei Fu, Jianfeng Feng, Tao Xiang, Philip~HS Torr, et~al.
\newblock Rethinking semantic segmentation from a sequence-to-sequence
  perspective with transformers.
\newblock In {\em Proceedings of the IEEE/CVF conference on computer vision and
  pattern recognition}, pages 6881--6890, 2021.

\bibitem{lin2022ds}
Ailiang Lin, Bingzhi Chen, Jiayu Xu, Zheng Zhang, Guangming Lu, and David
  Zhang.
\newblock Ds-transunet: Dual swin transformer u-net for medical image
  segmentation.
\newblock {\em IEEE Transactions on Instrumentation and Measurement}, 2022.

\bibitem{tan2021analysis}
Wenjun Tan, Peifang Huang, Xiaoshuo Li, Genqiang Ren, Yufei Chen, and Jinzhu
  Yang.
\newblock Analysis of segmentation of lung parenchyma based on deep learning
  methods.
\newblock {\em Journal of X-ray science and technology}, 29(6):945--959, 2021.

\bibitem{tan2021automated}
Wenjun Tan, Luyu Zhou, Xiaoshuo Li, Xiaoyu Yang, Yufei Chen, and Jinzhu Yang.
\newblock Automated vessel segmentation in lung ct and cta images via deep
  neural networks.
\newblock {\em Journal of X-Ray Science and Technology}, (Preprint):1--15,
  2021.

\bibitem{tan2022segmentation}
Wenjun Tan, Pan Liu, Xiaoshuo Li, Shaoxun Xu, Yufei Chen, and Jinzhu Yang.
\newblock Segmentation of lung airways based on deep learning methods.
\newblock {\em IET Image Processing}, 16(5):1444--1456, 2022.

\bibitem{armato2011lung}
Samuel~G Armato~III, Geoffrey McLennan, Luc Bidaut, Michael~F McNitt-Gray,
  Charles~R Meyer, Anthony~P Reeves, Binsheng Zhao, Denise~R Aberle, Claudia~I
  Henschke, Eric~A Hoffman, et~al.
\newblock The lung image database consortium (lidc) and image database resource
  initiative (idri): a completed reference database of lung nodules on ct
  scans.
\newblock {\em Medical physics}, 38(2):915--931, 2011.

\bibitem{qin2019airwaynet}
Yulei Qin, Mingjian Chen, Hao Zheng, Yun Gu, Mali Shen, Jie Yang, Xiaolin
  Huang, Yue-Min Zhu, and Guang-Zhong Yang.
\newblock Airwaynet: a voxel-connectivity aware approach for accurate airway
  segmentation using convolutional neural networks.
\newblock In {\em International Conference on Medical Image Computing and
  Computer-Assisted Intervention}, pages 212--220. Springer, 2019.

\bibitem{yu2022break}
Weihao Yu, Hao Zheng, Minghui Zhang, Hanxiao Zhang, Jiayuan Sun, and Jie Yang.
\newblock Break: Bronchi reconstruction by geodesic transformation and skeleton
  embedding.
\newblock In {\em 2022 IEEE 19th International Symposium on Biomedical Imaging
  (ISBI)}, pages 1--5. IEEE, 2022.

\bibitem{zhang2021fda}
Minghui Zhang, Xin Yu, Hanxiao Zhang, Hao Zheng, Weihao Yu, Hong Pan, Xiangran
  Cai, and Yun Gu.
\newblock Fda: Feature decomposition and aggregation for robust airway
  segmentation.
\newblock In {\em Domain Adaptation and Representation Transfer, and Affordable
  Healthcare and AI for Resource Diverse Global Health}, pages 25--34.
  Springer, 2021.

\bibitem{cciccek20163d}
{\"O}zg{\"u}n {\c{C}}i{\c{c}}ek, Ahmed Abdulkadir, Soeren~S Lienkamp, Thomas
  Brox, and Olaf Ronneberger.
\newblock 3d u-net: learning dense volumetric segmentation from sparse
  annotation.
\newblock In {\em International conference on medical image computing and
  computer-assisted intervention}, pages 424--432. Springer, 2016.

\bibitem{hofmanninger2020automatic}
Johannes Hofmanninger, Forian Prayer, Jeanny Pan, Sebastian R{\"o}hrich, Helmut
  Prosch, and Georg Langs.
\newblock Automatic lung segmentation in routine imaging is primarily a data
  diversity problem, not a methodology problem.
\newblock {\em European Radiology Experimental}, 4(1):1--13, 2020.

\bibitem{milletari2016v}
Fausto Milletari, Nassir Navab, and Seyed-Ahmad Ahmadi.
\newblock V-net: Fully convolutional neural networks for volumetric medical
  image segmentation.
\newblock In {\em 2016 fourth international conference on 3D vision (3DV)},
  pages 565--571. IEEE, 2016.

\bibitem{lin2017focal}
Tsung-Yi Lin, Priya Goyal, Ross Girshick, Kaiming He, and Piotr Doll{\'a}r.
\newblock Focal loss for dense object detection.
\newblock In {\em Proceedings of the IEEE international conference on computer
  vision}, pages 2980--2988, 2017.

\bibitem{isensee2021nnu}
Fabian Isensee, Paul~F Jaeger, Simon~AA Kohl, Jens Petersen, and Klaus~H
  Maier-Hein.
\newblock nnu-net: a self-configuring method for deep learning-based biomedical
  image segmentation.
\newblock {\em Nature methods}, 18(2):203--211, 2021.

\bibitem{irving20093d}
Benjamin Irving, Paul Taylor, and Andrew Todd-Pokropek.
\newblock 3d segmentation of the airway tree using a morphology based method.
\newblock In {\em Proceedings of 2nd international workshop on pulmonary image
  analysis}, pages 297--07, 2009.

\bibitem{feuerstein2009adaptive}
Marco Feuerstein, Takayuki Kitasaka, and Kensaku Mori.
\newblock Adaptive branch tracing and image sharpening for airway tree
  extraction in 3-d chest ct.
\newblock In {\em Proc. of Second International Workshop on Pulmonary Image
  Analysis}, volume~1, pages 1--8, 2009.

\bibitem{inoue2013robust}
Tsutomu Inoue, Yoshiro Kitamura, Yuanzhong Li, and Wataru Ito.
\newblock Robust airway extraction based on machine learning and minimum
  spanning tree.
\newblock In {\em Medical Imaging 2013: Computer-Aided Diagnosis}, volume 8670,
  pages 141--149. SPIE, 2013.

\bibitem{xu2015hybrid}
Ziyue Xu, Ulas Bagci, Brent Foster, Awais Mansoor, Jayaram~K Udupa, and
  Daniel~J Mollura.
\newblock A hybrid method for airway segmentation and automated measurement of
  bronchial wall thickness on ct.
\newblock {\em Medical image analysis}, 24(1):1--17, 2015.

\bibitem{yun2019improvement}
Jihye Yun, Jinkon Park, Donghoon Yu, Jaeyoun Yi, Minho Lee, Hee~Jun Park,
  June-Goo Lee, Joon~Beom Seo, and Namkug Kim.
\newblock Improvement of fully automated airway segmentation on volumetric
  computed tomographic images using a 2.5 dimensional convolutional neural net.
\newblock {\em Medical image analysis}, 51:13--20, 2019.

\bibitem{crisafulli2017prevalence}
Ernesto Crisafulli, Roberta Pisi, Marina Aiello, Matteo Vigna, Panagiota Tzani,
  Anna Torres, Giuseppina Bertorelli, and Alfredo Chetta.
\newblock Prevalence of small-airway dysfunction among copd patients with
  different gold stages and its role in the impact of disease.
\newblock {\em Respiration}, 93(1):32--41, 2017.

\bibitem{xiong2020clinical}
Ying Xiong, Dong Sun, Yao Liu, Yanqing Fan, Lingyun Zhao, Xiaoming Li, and
  Wenzhen Zhu.
\newblock Clinical and high-resolution ct features of the covid-19 infection:
  comparison of the initial and follow-up changes.
\newblock {\em Investigative radiology}, 2020.

\end{thebibliography}






\end{document}